%% file: StructureDynamicsStatsReview_Final_Upload.tex
\documentclass[review,onefignum,onetabnum]{siamart171218}

\input{shared}

\usepackage{color}
\usepackage{enumitem}
\usepackage{todonotes}
\usepackage{cite}
\usepackage{mathrsfs}
\usepackage{lineno}

\newcommand{\set}[2]{\left\{\left.#1\,\right|\,#2\right\}}
\newcommand{\Ps}{\mathscr{P}}
\newcommand{\maxdim}{N}

\renewcommand{\top}{\mathsf{T}}

\newcommand{\eps}{\varepsilon}
\newcommand{\C}{\mathbb{C}}
\newcommand{\Z}{\mathbb{Z}}

\newcommand{\Tor}{\mathbb{T}}

\newcommand{\bm}[1]{\mathbf{#1}}

\newtheorem{thm}{Theorem}[section]

\theoremstyle{definition}
\newtheorem{defn}[thm]{Definition}
\theoremstyle{remark}
\newtheorem{rem}[thm]{Remark}
\newtheorem{ex}[thm]{Example}

\usepackage{graphicx,amssymb,epsf,amsmath,setspace,float,dsfont, soul, tikz}
\usetikzlibrary{shadows,arrows,trees,shapes,matrix,decorations.pathmorphing,positioning}
\usetikzlibrary{patterns}
\usetikzlibrary{decorations.markings}
\usetikzlibrary{cd}
\usetikzlibrary{backgrounds,calc}
\pgfdeclarelayer{background}
\pgfdeclarelayer{foreground}
\pgfsetlayers{background,main,foreground}

\usetikzlibrary{decorations.markings,arrows.meta,bending}
\tikzset{->-/.style={decoration={markings, 
  mark=at position #1 with {\arrow[line width=.5pt]{>}}},postaction={decorate}}}

\newcommand{\R}{\mathbb{R}}

\newcommand{\Rd}{\R^d}
\newcommand{\F}{\mathbb{F}}
\DeclareMathOperator{\ima}{im}

\ifpdf
\hypersetup{
  pdftitle={What are higher-order networks?},
  pdfauthor={Christian Bick, Elizabeth Gross, Heather~A.~Harrington, and Michael~T.~Schaub}
}
\fi

\myexternaldocument{supplement}

\begin{document}



\maketitle
\begin{abstract}
{Network-based modeling of complex systems} and data using the language of graphs has become an essential topic across a range of different disciplines.
Arguably, this {graph}-based perspective derives {its} success from the relative simplicity of graphs: A graph consists of nothing more than a set of vertices and a set of edges, describing relationships between \emph{pairs} of such vertices.
This simple combinatorial structure makes graphs interpretable and flexible modeling tools.
The simplicity of graphs as system models, however, has been scrutinized in the literature recently. 
Specifically, it has been argued from a variety of different angles that there is a need for  \emph{higher-order networks}, which go beyond the paradigm of modeling pairwise relationships, as encapsulated by graphs.
In this survey article we take stock of these recent developments.
Our goals are to clarify (i)~what higher-order networks are, (ii)~why these are interesting objects of study, and (iii)~how they can be used in applications.
\end{abstract}


\section{Introduction}

{Over the last decades, there has been} a surge of interest in networks and network dynamical systems, which consist of interconnected entities, {to understand} a variety of complex systems. Examples range from biological systems, such as gene regulatory networks, to infrastructure systems, such as transportation networks~\cite{Strogatz2004,Barabasi2016,newman2018networks}.

{The term \emph{network} often refers to a \emph{graph}, which is a combinatorial structure that consists of vertices (or nodes) and edges (or links).
In this abstraction the nodes represent the entities in a system and the edges denote which entities interact.}
Representing systems as graphs has been instrumental to gain insights about the structural and dynamical features of a system:
Graph properties can be used to determine important nodes~\cite{friedkin1991theoretical,page1999pagerank,gleich2015pagerank}, reveal modular structure of a system~\cite{fortunato2016community,schaub2017many}, or---if each node is a dynamical unit---elucidate collective network dynamics such as synchronization~\cite{Strogatz2000}.
These ideas have been extended to weighted graphs, signed graphs, directed graphs, and graphs with multiple edge or node types such as multilayer or multiplex graphs.
However, a limitation of any graph-based approach is that all relationships are by definition \emph{dyadic} or \emph{pairwise relationships}, since edges in a graph are pairwise relations.
For instance, in network dynamics, dyadic interactions are typically reflected in the equation of motion {by the fact that the interaction between two given nodes are independent of any other node in the network}.

In many real-world systems, however, {network} interactions and relations are non-dyadic, and involve the joint nonlinear interaction of more than two nodes.
For instance, in socio-economic systems, where group-based interactions are commonplace, activity is often jointly coordinated between multiple agents (e.g., buyer, seller, broker)~\cite{bonacich2004hyper}.
Reactions in biochemical systems often involve more than two species (${A\!+\!B \rightarrow C\!+\!D}$)~\cite{klamt2009hypergraphs}, or two reagents might interact only in the presence of an enzyme.
The importance of nonlinear interactions between three or more nodes has also been long debated in the social sciences~\cite{easley2010networks}.
For instance, structural balance theory implies that triadic relations in social networks will evolve according to the colloquial rules ``the friend of a friend is my friend'' and ``the enemy of my enemy is my friend''~\cite{marvel2011continuous}.
Recent analyses of large online social networks verify that these networks are indeed extremely balanced~\cite{facchetti2011computing}.
Similarly, peer pressure, {the formation of coalitions and alliances, and brokerage activities are examples for common social phenomena, which do involve the intertwined activity of multiple people, rather than pairs; see~\cite{Battiston2020} for discussions on further examples and references.}
It has also been argued that joint interactions are crucial to see competition patterns in multiple interacting species~\cite{Abrams1983, Levine2017}.

To adequately capture the properties of any such {network}, it is then crucial to go beyond graphs, which only capture dyadic relationships, and to elucidate the effect of polyadic network interactions~\cite{benson2021higher,torres2020representations,Battiston2020}.
Polyadic interactions and relations have been discussed under a variety of names in the literature, including \emph{supra-dyadic}, \emph{nonpairwise}, \emph{higher-order}, or \emph{simplicial}.
In the following, we will collectively refer to networks with such interactions as \emph{higher-order networks}.
But what exactly are higher-order networks? How can they be represented mathematically? And what are the tools available to analyze them? 

One reason underlying the differences in nomenclature is that higher-order networks arise in various contexts throughout different research areas, which have often somewhat different motivations, research questions, and mathematical tools.
The main aim of this article is to provide an integrated perspective on higher-order networks and their mathematical representations, and review recent progress in this very active field of research, focusing on the underlying questions and mathematical tools.
Consequently, our review complements other recent surveys, including a more physics-centered perspective in~\cite{Battiston2020} that features an extensive list of references,
a discussion of relationships between different higher-order models and dependency modeling~\cite{torres2020representations}, and a survey on signal-processing on higher-order networks~\cite{Schaub2021}.
Specifically, we will discuss higher-order networks in three contexts, where they have found particular application:
\begin{itemize}
    \item \textbf{Topology and geometry of data} (\Cref{sec:Structure}): How do higher-order networks help in understanding the ``shape'' of datasets?
    \item \textbf{Analysing and modelling relational data} (\Cref{sec:Statistics}): How can higher-order relations be modeled statistically and what are advantages of using  
    {such statistical} models?
    \item \textbf{Higher-order network dynamical systems} (\Cref{sec:Dynamics}): How can higher-order interactions effect the collective dynamics of coupled dynamical units?    
\end{itemize}
We believe that this overview of higher-order networks can provide a common language for a number of research areas that have predominantly been considered separately.
We highlight differences and similarities between those areas and formulate long term research perspectives.

\subsection{Beyond dyadic interactions: From graphs to higher-order networks}
\label{sec:Preview}

\newcommand{\G}{\mathcal{G}} 	
\newcommand{\Hg}{\mathcal{H}}	
\newcommand{\K}{\mathcal{K}}	
\newcommand{\Si}{\K}			
\newcommand{\Ve}{\mathcal{V}}	
\newcommand{\Ed}{\mathcal{E}}	

\newcommand{\sset}[1]{\left\{#1\right\}}	

Before diving into a more detailed discussion, we will preview the three topics outlined above.
To this end, we need to fix some terminology that will be used throughout the paper.
Mathematically more precise definitions of these objects will be given in~\Cref{sec:Prelims}.
An (undirected) graph~$\G$ consists of a finite set of vertices~$\Ve$ and edges~$\Ed$. 
An edge $e = \sset{v_k, v_j}\in \Ed$ indicates a relationship between the vertices~$v_k, v_j\in\Ve$.
A hypergraph~$\Hg$  on the set of vertices~$\Ve$ is a generalization of a graph: The hyperedges $e\in\Ed$ are arbitrary non-empty subsets $e\subseteq\Ve$.
An abstract simplicial complex~$\Si$ is a hypergraph~$\Hg$ such that the set of hyperedges is closed under inclusion, that is, if $d\subset e\in\Ed$ and $d\neq\emptyset$ then $d\in\Ed$.
Abstract simplicial complexes are combinatorial objects that relate to their geometric counterparts in classical algebraic topology~\cite{Hatcher2002}.  
Hypergraphs and simplicial complexes will play a main role as we discuss higher-order networks in this review.

\subsubsection{Topology and geometry: Understanding the ``shape'' of data}\label{ssec:topology_preview}

\begin{figure}
    \centering
    \includegraphics[width=\textwidth]{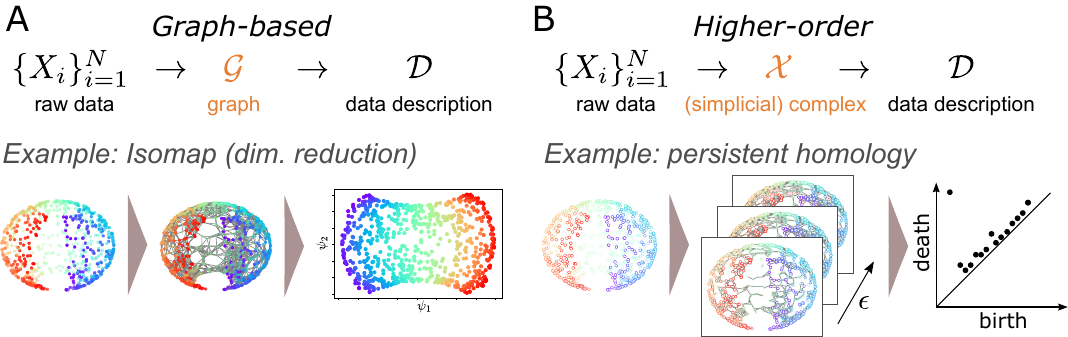}
    \caption{\textbf{Schematic: Higher-order networks to describe geometry and topology of data.} 
        \textbf{A} The Isomap algorithm~\cite{tenenbaum2000global} is an instance of a manifold learning algorithm~\cite{ma2011manifold}, operating as follows.
        We associate each point to a vertex of a graph. Two such vertices are connected by an edge if the corresponding nodes are within a certain distance $\epsilon$ from each other.
        By analyzing the resulting graph, we can define a new set of distances between the points which can be used to assign a new set of (low dimensional) coordinates to each point, which reveal the underlying geometric ``shape'' of the data.
{                \textbf{B} Persistent homology avoids selecting a specific threshold value for $\epsilon$ by analyzing data at multiple scales~\cite{edelsbrunner2008,Zomorodian2005}. 
        The persistent homology algorithm first builds a nested sequence of spaces on data, called a filtered simplicial complex, which is indexed by a scale parameter (e.g., $\epsilon$). An invariant in algebraic topology, homology, gives a way to measure topological information, such as the number of connected components, loops or voids. Persistent homology of the filtration outputs a multiset of intervals, where the left end points correspond to the scale value at which a specific feature topological feature appears, the right end point corresponds to the value it disappears, and the length of the interval is the persistence of such a feature. This topological summary can be visualised as a persistence diagram or barcode~\cite{edelsbrunner2008,Zomorodian2005}. In this way various higher-order connectivity aspects of the dataset can be assessed (see \Cref{sec:Structure}).
        }
}%
        \label{fig:schematic_higher_order_geometric}
\end{figure}

\newcommand{\sdim}{n}

High-dimensional point-cloud data commonly contains some geometric or topological information. 
Topological tools enable the classification of \emph{topological or geometric structure}, for instance, the detection of possible clusters in the data. 
As an example, consider the Isomap algorithm~\cite{tenenbaum2000global} illustrated in~\Cref{fig:schematic_higher_order_geometric}A, that aims to learn a low-dimensional representation of point cloud data. 
The central idea of Isomap is that if the observed data is assumed to be a noisy sample from a low-dimensional manifold embedded in a larger sample space, we can construct a graph that serves as a discrete proxy for this continuous object: The vertices correspond to the data points, and two vertices are connected by an edge if the distance between the corresponding data points is smaller than some predefined parameter~$\eps$.
The resulting (family of) graphs~$\G_\eps$ provides a low-dimensional representation of the data. A number of popular algorithms follow a variation of this paradigm: (i)~map each data point to a vertex in a graph, (ii)~define edges according to some criteria based on \emph{pairwise} similarity or distance of the original data points, and (iii)~analyze the constructed graph to extract a low-dimensional representation of the nodes, thereby providing an embedding of the original data points associated to those nodes.
For instance, diffusion maps~\cite{coifman2006diffusion}, or Laplacian eigenmaps~\cite{belkin2003laplacian}, define embedding coordinates based on the (scaled) eigenvectors of a Laplacian matrix that is derived from a graph  $\G_\eps$ constructed in the way outlined above.
For an overview of these kind of manifold learning algorithms, see also the book~\cite{ma2011manifold} and references therein.

The idea of approximating the global geometric structure of high-dimensional data using graphs naturally generalizes beyond local pairwise geometric relationships to local $\sdim$-wise geometric relationships (cf.~\Cref{sec:Structure}).
For instance, analogous to the graph construction based on a distance threshold~$\eps$, we may construct a \v{C}ech complex~$\K_\eps$ as a family of simplicial complexes: Place a ball of radius~$\eps$ around each data point and declare that there is a $(\sdim-1)$-simplex between any~$\sdim$ points if the intersection of the corresponding $\eps$-balls is non-empty. 
Clearly such a simplicial complex based description encapsulates more information on the geometry of the data compared to a graph  (see ~\Cref{fig:schematic_higher_order_geometric}B).
Consequently, the higher-order network given by a simplicial complex contains more information about the data: For instance, simplicial complexes can be used to reveal certain aspects of topology of the data by means of a technique called persistent homology~\cite{edelsbrunner2008,Zomorodian2005}.
Simplicial complexes also give rise to a hierarchy of Hodge-Laplacians~\cite{lim2020hodge,Schaub2020}, which include the graph Laplacian matrix as a special case, and can be used to extract geometric and topological information about the data.

\subsubsection{Modeling relational data via higher-order networks}\label{ssec:relational_data_preview}    
\begin{figure}
    \centering
    \includegraphics[]{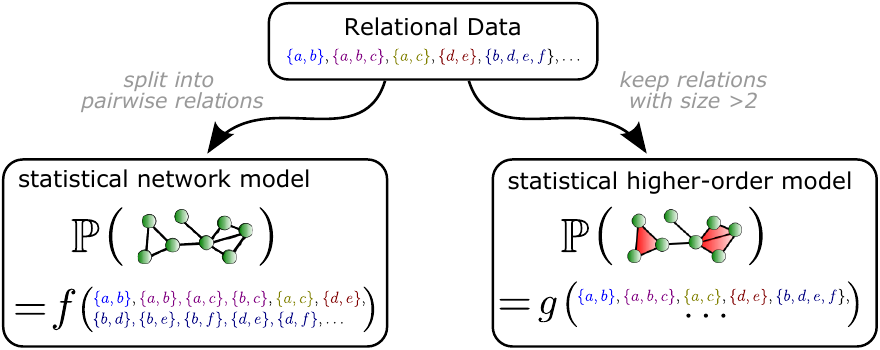}
    \caption{\textbf{Schematic: Higher-order networks to describe relational data.}
        Given some empirically observed relationships between sets of entities (top), e.g., co-authorship relations between scientists, a typical approach to analyze such data is to abstract these relationships as the union of pairwise relationships, representing the edges in a graph. In other words, we decompose the observed relationships into pairwise relations and define a graph based on those pairwise relations. Our task is  typically to analyze this graph by means of specifying a generative model for the observed data (bottom left).
        Rather than settling on a graph representation, we can alternatively try to keep all those (potentially) higher-order relationships intact and represent and analyze them directly in terms of a hypergraph or simplicial complex (bottom right). 
        For instance, we again try to specify a probabilistic model for the empirically observed data.
    }%
        \label{fig:schematic_higher_order_rel_data}
\end{figure}

The statistical analysis of networks, such as friendship relationships in a social network, has been a mainstay topic of network analysis.
The data we would like to understand in this context is typically \emph{relational data}, i.e., data about how \emph{sets of entities} are related (see~\cref{fig:schematic_higher_order_rel_data}).
Further examples include the modeling and description of co-authorship networks,  ecological relationships (predator-prey or mutualistic), or hyperlink graphs describing the world-wide-web, to name a few.
In contrast to the toplogical/geometrical or the dynamical perspective discussed in \Cref{ssec:topology_preview,ssec:dynamics_preview}, when adopting this perspective we are primarily concerned with statistical modeling of the relations themselves, rather than analyzing those relations to learn about some topological and geometrical shape of the network.
Put differently, our goal is to provide a statistical model that specifies a probability of observing a set of given relations.

In many applications, relationships are recorded between exactly two entities---such as hyperlinks from one webpage to another, or a friendship connection between two people---naturally giving rise to edges in a (dyadic) graph linking the two entities, where the entities are represented as vertices.
Typical models for relational data thus define probability distributions over a set of graphs.
A simple example is the Erd\H os-R\'enyi model that posits that for a given a set of vertices $\Ve$ of cardinality~$\maxdim$, there exists a pairwise relation between any two objects with a fixed probability~$p$.  Other examples include configuration models, stochastic block models, or exponential random graph models.

However, as the example of co-authorships mentioned above illustrates, relational data may comprise sets with more than two entities: A paper may have more than two authors that appear jointly.
Rather than breaking such a relation between multiple entities up into a number of dyadic relationships, we may try to keep those relationship sets intact, and directly take them into account in our statistical models.  
It then becomes necessary to develop probabilistic models over spaces of hypergraphs or simplicial complexes.  We discuss and review such statistical models, and the questions that arise in their development and analysis in~\Cref{sec:Statistics}.

\subsubsection{Higher-order network dynamical systems}\label{ssec:dynamics_preview}

\begin{figure}[tb!]
    \centering
    \includegraphics[]{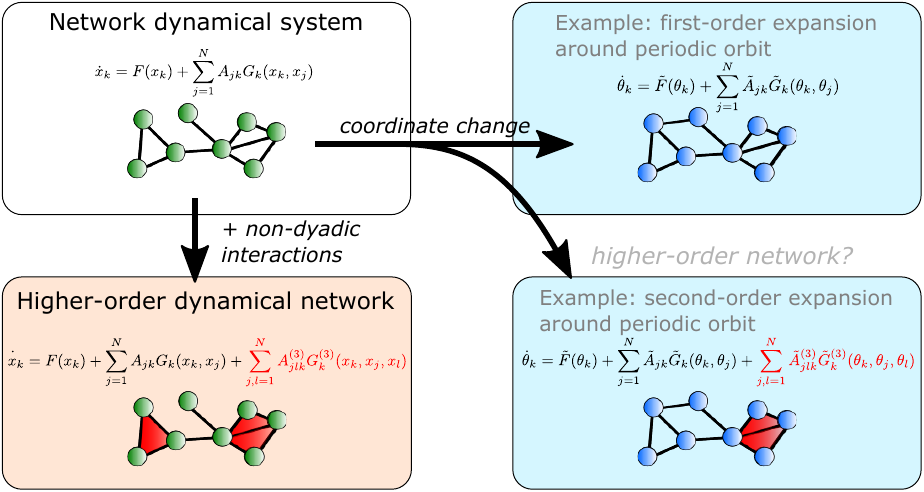}
    \caption{\textbf{Schematic: Higher-order networks to describe couplings between dynamical units.}
        {Dynamics on graphs determined by ordinary differential equations are examples of network dynamical systems: Each node has intrinsic dynamics and nodes interact if they are adjacent (top left).}
        A natural extension {dynamics on a hypergraph, which allow for multi-way interactions between more than two nodes}, such as triplets of nodes, etc.~(bottom left).
        However, care is needed as a change of coordinates of the original dyadic network dynamical system, may also result in an ``apparent'' higher-order dynamical system, depending on the type of coordinate transformation used (top right vs bottom right).
}%
        \label{fig:schematic_higher_order_dynamic}
\end{figure}

Network dynamical systems provide a natural abstraction for many real-world systems that consist of interacting dynamical entities. 
These include networks of coupled neurons, power grid networks, ecological networks of interacting species, and epidemic processes on networks~\cite{boccaletti2006complex}. 
A network dynamical system typically consists of (i)~a number of dynamical nodes with state variables that evolve according to 
differential equations and (ii)~a network structure that captures which nodes interact and how they interact (the functional form). The principal question is: How does the collective network dynamics---the joint dynamics of all nodes, such as synchronization---relate to the network structure?

In traditional network dynamical systems, the interactions between nodes are captured by a graph; cf.~\Cref{fig:schematic_higher_order_dynamic}. 
The graph may be based on actual physical connections (e.g., a physical power line between two nodes), imposed as a modeling choice (e.g., a dynamical system on a graph), inferred from data (e.g., an edge is placed between two nodes when the time series are correlated), or given as a formal description of dependencies between node states (e.g., in coupled cell systems).
Moreover, the network interactions are often additive: For a network of Kuramoto oscillators~\cite{Kuramoto,Acebron2005,Rodrigues2016}---one of the most prominent examples of network dynamical system---the joint effect of two oscillators on a third oscillator is the sum of the two individual (nonlinear) effects.
Synchronization occurs when different nodes (eventually) behave in unison as time evolves. 
Such dynamical properties are influenced by the properties of the graph that represents the network. 
For example, classical results relate the spectrum of the underlying graph to the synchronization properties of the network dynamical system~\cite{Pecora1998}. 

{A natural generalization of network dynamical systems is to allow joint nonlinear interactions} between three or more oscillators---this results in a \emph{higher-order network dynamical system}. 
Such a generalization, however, raises the following questions: What is an appropriate combinatorial structure to capture such interactions? How do dynamical properties, such as synchronization, relate to the algebraic properties of this structure? One approach is to define a dynamical system on a hypergraph or simplicial complex, which has certain caveats. As we will discuss in~\Cref{sec:Dynamics}, for coupled oscillator networks, a coordinate transformation may also yield ``effective'' nondyadic interactions for systems that are additively coupled; see~\Cref{fig:schematic_higher_order_dynamic}. 
Indeed, a (higher-order) network representation of a dynamical system is not necessarily invariant under coordinate transformations. 
More importantly, however, if a higher-order network dynamical system is related to a dyadic network dynamical system by a coordinate transformation, it will exhibit the same dynamical properties.
Stated differently, if there exists such a transformation, then the overall system can equivalently be described by an effective (dyadic) network and no higher-order network representation is necessary.
We will discuss these aspects and their implications for modeling in more detail in~\Cref{sec:Dynamics}.

\subsubsection{An Example: The many facets of higher-order interactions}
\label{sec:Celegans}
Let us highlight the difference in the above discussed perspectives by considering the example of \emph{C.~elegans}, a roundworm whose complete connectome---the physical connections between neurons---has been mapped out.
Importantly, the connectome is the same for each of these worms and thus there is hope that analysing the connectome in the form of a (higher-order) network  may tell us about important features of the neural information processing within the worm.
Now, there are several types of questions we can ask.
First, we may be interested in the topological features of this connectome by using the techniques in \Cref{sec:Structure} to analyze the ``shape'' of this network~\cite{sizemore2019importance}.
For instance, persistent homology gives insights into the topological features of this specific connectome.
Second, 
we may be interested in specifying a probabilistic network model of the connectomic data (\Cref{sec:Statistics}). Even though the connectome of every worm is the same and there is thus nothing stochastic about the connectome of \emph{C.~elegans}, treating the connections as relational data to which we want to fit a probabilistic model may provide us with certain insights{, e.g., whether or not certain connection patterns are expressed more or less often than would be expected.}
For instance, by comparing to a random null model, we can assess how likely it is that certain features of the connectome have arisen by chance~\cite{yan2017network, gross2021random}. 
Third, 
network properties can lead to an understanding of the collective dynamics of \emph{C.~elegans}' neuronal network~\cite{varshney2011structural,bacik2016flow,yan2017network}. While one may intuitively think of the connectome as a graph, both nonadditive network coupling and effective (or indirect) interactions can give rise to a higher-order network dynamical system~(\Cref{sec:Dynamics}).

\subsection{Missing links: Topics not covered}\label{sec:relatedButNotCovered}
There are some other topics that can intuitively be considered from the perspective of higher-order networks but are beyond the scope of this review. 
We give a brief overview and provide references for further reading.

\paragraph{Multiplex networks}
Multiplex, multilayer, and networks of networks have been proposed as modeling paradigms for systems where there are different types of interactions~\cite{Kivelae2014}. These aim to account for links of different types, for example, different modalities (phone, text, e-mail) in a communication network. In most cases, however, the interactions are dyadic and can thus be presented by traditional networks.

\paragraph{Higher-order Markov models for sequential data}
Markov models defined on networks have become a popular way to describe and model flows of information, energy, mass, money, etc.~between different entities.
If the evolution is given by a (first-order) Markov process, the process can be seen as a random walk on a graph~\cite{masuda2017random}. However, many empirically observed flows on networks do have some path dependency. Thus higher-order Markov chain models are required~\cite{Lambiotte2019}.

\paragraph{Higher-order graphical models and Markov random fields}
Markov random fields such as the Ising model and more general graphical models
have also been extended to higher-order models that account for interactions between multiple entities. 
For instance, for the Ising model, extensions have been sought that include interactions of third order to model social interactions in mice~\cite{shemesh2013high}.
Other works about incorporating higher-order interactions in graphical models include~\cite{zheleva2010higher,pirayre2017hogmep,komodakis2009beyond}.
{Note that, a graphical model can of course be used to describe probability distributions over various types of objects, including graphs and the  generalizations of graphs discussed in~\Cref{sec:Statistics}.
However, the emphasis in~\Cref{sec:Statistics} lies on statistical models for dyadic and non-dyadic \emph{relational data}, rather than on higher-order dependencies built into a statistical model, e.g., in terms of third-order interactions in an extended Ising model~\cite{shemesh2013high}.}

\paragraph{Graph signal processing and simplicial signal processing}
Graph signal processing is an relatively new area of signal processing that deals with the processing of signals supported on the nodes of a graph, and it aims to translate and extend signal processing techniques such as interpolation, signal smoothing, sampling and filtering to the domain of graphs~\cite{shuman2013emerging,ortega2018graph}.
Recently, there have been propositions to extend these ideas to simplical complexes and hypergraphs; see, e.g.,~\cite{zhang2019introducing,barbarossa2020topological,Schaub2021,schaub2022signal}.

\paragraph{{Tensor based models, tensor based algorithms and applications}}
As (dyadic) networks can be described by matrices such as the adjacency matrix or the Laplacian, tensors provide another possible direction in which network models can be extended.
{For instance, we may consider generative processes of higher-order networks that are based on tensor quantities. Alternatively, we may use tensor-based algorithms such as tensor decompositions to analyze higher-order network data.}
Examples include the application of tensor {decompositions} for the detection of communities in time-dependent networks~\cite{gauvin2014detecting}, or spectral clustering of motifs~\cite{benson2015tensor}, as well as centrality analysis based on tensors~\cite{gleich2015multilinear}.

\subsection{Outline of this article}
The remainder of the article is structured as follows. 
In \Cref{sec:Prelims}, we collect the mathematical concepts that will be used throughout this paper. 
In \Cref{sec:Structure}, we discuss how to analyze topology and geometry of data, and we highlight how higher-order networks can help in understanding the structural features of datasets.
{In \Cref{sec:Statistics}, we discuss models of higher-order relations and how they can be useful in describing and understanding relational data.
In \Cref{sec:Dynamics}, we discuss higher-order network dynamical systems and their relationship with traditional network dynamical systems.}
We conclude with a discussion and an outline of questions for further research in \Cref{sec:discussion}.

\section{A brief review of graphs, hypergraphs, and simplicial complexes}
\label{sec:Prelims}

{Hypergraphs and simplicial complexes are mathematical objects that can encode higher-order interactions. }
Building on the informal introduction in \Cref{sec:Preview}, here we collect the definitions of these objects.

\subsection{Graphs}

{An undirected \emph{graph}~$\G = (\Ve,\Ed)$ consists of (i)~a finite set ${\Ve = \sset{v_1,\dotsc, v_\maxdim}}$  of~$\maxdim$ vertices and (ii)~a set~$\Ed \subseteq \set{ \sset{ v_i,v_j}}{v_i,v_j\in \Ve}$ of edges, corresponding to \emph{unordered} pairs of elements of~$\Ve$. 
In contrast, a \emph{directed graph} $\G$ is a pair $\G = (\Ve, \Ed)$ of a finite  set of vertices~$\Ve = \sset{v_1,\dotsc,v_\maxdim}$ and a set of edges (or arrows) $\Ed \subseteq \Ve\times \Ve$ consisting of \emph{ordered} pairs of vertices. 
For ease of notation, we will typically fix a numbering of the vertices and set $\Ve=\sset{1,\dotsc,\maxdim}.$
}

{A graph is called \emph{simple} if it does not have self-loops, i.e., in a simple graph a vertex cannot be connected to itself by an edge.
There are notions of graphs more general than (simple) graphs. 
For example, in a \emph{multigraph} we can multiple different edges between the same nodes, rather than just one.}

{For a directed edge $(t,h) \in \Ed$ we say that it has \emph{head}~$h$ and \emph{tail}~$t$; edge~$e$ is a \emph{self-loop} if $h=t$.
Any graph~$\G$ with vertices $\Ve=\sset{1,\dotsc,\maxdim}$ can be identified with an~$\maxdim\times\maxdim$ adjacency matrix~$\bm A=(A_{kj})$ with $A_{kj} = 1$ if $(k,j)\in \Ed(\G)$ and $A_{kj}=0$ otherwise.
For an undirected graph~$\G$, the matrix~$\bm A$ is symmetric, i.e., $\bm{A}\!^\top =\bm A$ where~${}^\top$ denotes the transpose. }

{In many applications, it is useful to consider graphs with weighted edges.
A \emph{weighted graph} is a graph with $\G = (\Ve, \Ed)$ together with weights $w_{kj}\neq 0$ for each edge $\sset{v_k,v_j}\in\Ed$.
A weighted, directed graph is defined analogously.
For a weighted graph~$\G$ we define the \emph{weighted adjacency matrix} with~$A_{kj}=w_{kj}$ if $(v_k,v_j)\in \Ed(\G)$ and $A_{kj}=0$ otherwise.}

\subsubsection{Hypergraphs}
An undirected \emph{hypergraph} generalizes an undirected graph.
Let~$\Ps(X)$ denote the power set of a set~$X$, consisting of all subsets of~$X$.  

\begin{defn}[Hypergraph~\cite{Berge1973}]
    Let {$\Ve=\sset{1,\dotsc,\maxdim}$} be a finite set and $\Ed\subseteq \Ps(\Ve)$ a finite collection of non-empty subsets of~$\Ve$. 
    The tuple $\Hg = (\Ve, \Ed)$ is called a \emph{hypergraph}.
    A \emph{$k$-uniform hypergraph} is a hypergraph such that all hyperedges $e \in \Ed$ have cardinality $|e|=k$.
\end{defn}

\begin{figure}
    \centering
    \includegraphics[width=1.2in]{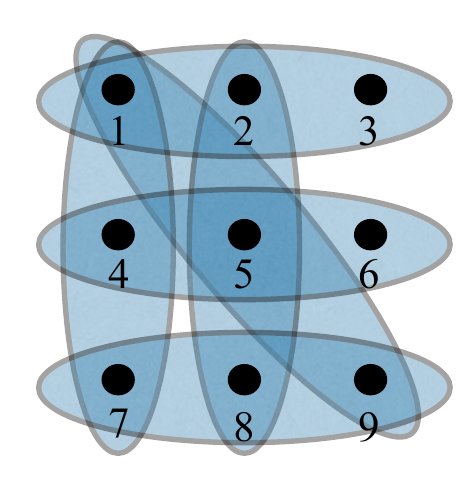}
    \caption{A 3-uniform hypergraph with 9 vertices and 6 edges.}%
        \label{fig:hypergraphexample}
\end{figure}

\begin{ex}  Let $\Ve = \{1, \ldots, 9\}$ and $\Ed = \{ \{1, 2, 3\}, \{4, 5, 6\}, \{7, 8, 9\},$ \\ $\{1, 4, 7\}, \{2, 5, 8\}, \{1, 5, 9\}\}$.  The hypergraph $\Hg = (\Ve, \Ed)$ is a 3-uniform hypergraph and is illustrated in \Cref{fig:hypergraphexample}.
\end{ex}

Hypergraphs, like graphs, can be extended to incorporate directionality and weights; see for example~\cite{Ausiello2017} and references therein.
In the context of higher-order networks in statistics, we can think of integer weights in terms of hypergraphs in which certain edges appear multiple times.

\subsubsection{Simplicial complexes}

An abstract simplicial complex is a hypergraph where the set of hyperedges is closed under inclusion. Specifically, we require that every non-empty subset~$d$ of an hyperedge $e\in\Ed$ is also part of the set of hyperedges.

\begin{defn}[Abstract simplicial complex]
{
    Consider a finite set of vertices $\Ve = \sset{1,\dotsc,\maxdim}$. A \emph{$\sdim$-simplex}
    $\sigma=\sset{v_0, \dotsc, v_\sdim}$ is a non-empty set of~$\sdim+1$ elements of~$\Ve$ and its dimension is defined to be~$\sdim$.
    Given an abstract $\sdim$-simplex~$\sigma$, we call a subset $\tau\subset\sigma$ a face of $\sigma$. 
    An \emph{abstract simplicial complex}~$\Si$ is a finite collection of such simplices such that 
\begin{itemize}
    \item $\tau\subset\sigma$ and $\sigma\in \Si$ imply $\tau\in \Si$ 
    \item $\{v\}\in \Si$ for all $v\in \Ve$. 
\end{itemize}
The \emph{dimension} of an abstract simplicial complex~$\Si$ is defined as the largest dimension of any simplex in~$\Si$.  The \emph{$\sdim$-skeleton}~$\Si_\sdim$ of the abstract simplicial complex~$\Si$ is the union of the simplices of dimensions ${0,1,\dotsc,\sdim}$.
The 0-skeleton~$\Si_0=\Ve$ are the vertices of~$\Si$, and the 1-skeleton is the graph associated to the simplicial complex. 
}
\end{defn}

{\rem{Different communities of researchers use different conventions for vertex numbering. For graphs and hypergraphs, vertices are often indexed starting with~$1$, whereas in algebraic topology, they start with~$0$. As mentioned above, we here number the vertices and set $\Ve=\sset{1,\dotsc,\maxdim}$ so that the maximal possible simplex dimension is~$N-1$; while not standard in computational algebraic topology, it eases notation. 
Note that we use $n$ to refer to a generic $n$-simplex of arbitrary dimension $n$ ($n+1$ vertices), whereas we use $N$ as a fixed constant describing the total number of vertices.
}}

\begin{rem}
To emphasize that simplicial complexes are a special type of hypergraph, one can use the notation $\Si=(\Ve, \Ed) =(\Si_0,\Ed)$, with~$\Ve$ the elements of~$\Si_0$ and~$\Ed$ being the set of all simplices contained in~$\Si$.
\end{rem}

\begin{ex}\label{example}
Given $\Ve=\sset{1,\dotsc,\maxdim}$, consider the abstract simplicial complex
\begin{equation}
\begin{split}
    \Si &= \{ \{1\}, \{2\},\{3\},\{4\}, \{5\},  \{1,2\}, \{1,3\},\{2,3\},\\
      &\quad\quad \{2,4\}, \{3,4\}, \{3,5\}, \{4,5\},  \{1, 2, 3\} \}.
\end{split}
\end{equation}
The simplicial complex~$\Si$ has dimension $2$ as $\{1, 2, 3\}$ is the largest simplex in $\Si$. 
There are five vertices, called 0-simplices: $\{1\}, \{2\},\{3\},\{4\}, \{5\}$; seven 1-simplices: $\{1,2\}, \{1,3\},\{2,3\}, \{2,4\},\{3,4\}, \{3,5\}, \{4,5\}$; and a single 2-simplex: $\{1, 2, 3\}$. 
This abstract simplicial complex can be visualized as follows.
\begin{center}
\scalebox{0.6}{
\begin{tikzpicture}
\newcommand*\pointsWPyul{332.9841032698848/400/0/1,415.8210384392016/241.07146003690823/1/2,507.4378311611366/400/2/3,595.7574064827379/239.23585016616542/3/4,680.7364875362073/400/4/5}
         \newcommand*\edgesWPyul{0/1,0/2,1/2,1/3,2/3,2/4,3/4}
\newcommand*\facesWPyul{0/1/2/10}
\newcommand*\scaleWPyul{0.02}          \foreach \x/\y/\z/\w in
\pointsWPyul {          \node (\z) at (\scaleWPyul*\x,-\scaleWPyul*\y)
[circle,draw,fill=white] {$\w$};          }          \foreach
\x/\y/\z/\w in \facesWPyul {    \fill[blue!25]    (\x.center) --
(\y.center) -- (\z.center) -- cycle;         }       \foreach
\x/\y/\z/\w in \pointsWPyul {          \node (\z) at
(\scaleWPyul*\x,-\scaleWPyul*\y) [circle,draw,fill=black!30] {$\w$};
    }             \foreach \x/\y in \edgesWPyul {          \draw (\x)
-- (\y);          } 
\end{tikzpicture}}
\end{center}
\end{ex}

{
To facilitate computations with simplicial complexes\footnote{Note that this in particular relevant when performing computations over the field $\mathbb{R}$, for binary coefficients this is not required.}, refer to simplices and their faces, or study directionality of simplices, we introduce an orientation for each simplex in~$\Si$.
Consider a $\sdim$-simplex $\sset{v_0,\dotsc,v_{\sdim}}$ that consists of an arbitrary set of $\sdim+1$ vertices $v_i\in\Ve$.
An \emph{oriented simplex} associated to this $\sdim$-simplex is denoted by an ordered tuple $\sigma=[v_0,\dotsc,v_{\sdim}]$.
There are two possible orientations: 
Two oriented simplicies associated to the same underlying simplex are equivalent if they differ simply by an even permutation of the tuple.
If two oriented simplices differ by an odd permutation they have an opposite orientation.
For instance, the oriented $1$-simplex (edge) $[v_i, v_j]$ from~$v_i$ to~$v_j$ has an opposite orientation to $[v_j, v_i]$. 
One can think of oriented simplicies as higher-order directed networks, which have gained increasing attention in the growing field of directed topology \cite{fajstrup2016directed,lutgehetmann2020computing,grigor2020path,chowdhury2018persistent,chaplin2021first} (note, however, that oriented graphs and directed (hyper)graphs are different, in general).
An oriented 1-simplex has a direction attached to it whereas an oriented 2-simplex has a sense of rotation attached to it.
To fix a reference orientation in this paper, we say that the oriented $\sdim$-simplex $[v_0,\ldots,v_{\sdim}]$ has a positive orientation, if the tuple is ordered equivalently to the standard ordering according to increasing vertex labels.
}

\section{Geometry and topology of data with (higher-order) networks}\label{sec:Structure}
As outlined in \Cref{ssec:topology_preview}, a central idea of topological data analysis is to describe empirically observed data, such as point clouds, by a topological object such as a simplicial complex. The properties of the topological object then yield a way to describe the data. 
The Euler characteristic~$\chi$ is an example of such a property, which 
gives a measure of size~\cite{schanuel1986length} in an appropriate sense. 
Similarly, \emph{homology}, which we will describe in the next sections, gives a way to measure  properties of simplicial complexes, such as the number of components and loops.
In fact homology is closely related to the Euler characteristic, but provides a more detailed account of the topology of an simplicial complex.
In the following, we first focus on homology as an important topological property of simplicial complexes, before we discuss how these concepts can be used to describe the higher-order structure of data. 
For more information on these topological concepts, see~\cite{Hatcher2002,munkres2018elements,alexandroff2012elementary,giblin2013graphs,matsumoto2002introduction,may1999concise,ghrist2014elementary,schenck2003computational}.

\subsection{Simplicial homology}
Topological properties of a simplicial complex~$\Si$ can be studied using (simplicial) homology. 
An \emph{$n$-chain} is a finite formal sum of oriented $n$-simplices $\sigma_i\in \Si$, written $c = \sum_i a_i \sigma_i$.
Here~$\sigma_i$ is an ordered $n$-simplex in~$\Si$ and~$a_i\in\F$ is a coefficient\footnote{Changing the coefficients can highlight specific features
and typical fields to consider are~$\F=\Z/2\Z$ for persistent homology computations (cf.~\Cref{sec:PersHom} below) or $\F=\R$ for Hodge--Laplacians (cf.~\Cref{sec:Hodge} below).
} in a field~$\F$. Note that there are only finitely many~$\sigma$ and thus only finitely many of the coefficients~$a_i$ are non-zero. 
Two $n$-chains are added component-wise, like polynomials, which depends on the field. 
{Changing the orientation of any simplex $\sigma_i$ corresponds to a change in sign of the corresponding coefficient.}

\begin{defn}[$\sdim$th chain group]
    The \emph{$\sdim$th-chain group}~$C_\sdim(\Si,\F)$ is a finitely generated vector space that is spanned by its oriented $\sdim$-simplices with coefficients in a field~$\F$. 
\end{defn}

The chain groups are finite vector spaces generated by the set of $\sdim$-simplicies of~$\Si$. 
The linear maps~$\partial_\sdim$ between vector spaces~$C_\sdim(\Si, \F)$ and~$C_{\sdim-1}(\Si,\F)$ are called \emph{boundary operators} or \emph{differentials}, which maps an $\sdim$-simplex~$\sigma=[v_0, \ldots, v_{\sdim}]$, spanned by vertices $v_0, \ldots, v_{n}$ to an alternating sum of its $(\sdim-1)$-dimensional faces obtained by omitting the~$j$th vertex:
\begin{equation}
\begin{split}
\partial_\sdim:\qquad C_\sdim(\Si,\F) &\rightarrow C_{\sdim-1}(\Si,\F),\\
  [v_0, \dotsc, v_{\sdim}]&\mapsto \sum_{j= 0}^{\sdim} (-1)^{j}
  [ v_0, \dotsc, v_{j-1}, v_{j+1}, \dotsc, v_{\sdim} ].
\end{split}
\end{equation}

\begin{ex}\label{ex:boundary}
For a 1-simplex, we compute the boundary: 
         \begin{equation*}
	 \begin{aligned}
                                \partial_1\left(
                               \raisebox{-0.26cm}{
\begin{tikzpicture}[bullet/.style={circle, fill=black!50, inner
sep=0pt, minimum width=0pt, label={[font=\bfseries]#1}},>=latex]
    \draw [fill=blue!25,->-/.list={2/3}] 
    (0,0) node[circle, draw, fill=black!50,
                        inner sep=0pt, minimum width=3pt, label=below:{$v_0$}] {}
    -- (1,0) node[circle, draw, fill=black!50,
                        inner sep=0pt, minimum width=3pt, label=below:{$v_1$}]  {};
\end{tikzpicture}}\right) &= 
                               \raisebox{-0.26cm}{
\begin{tikzpicture}[bullet/.style={circle, fill=black!50, inner
sep=0pt, minimum width=pt, label={[font=\bfseries]#1}},>=latex]
    \draw [fill=blue!25] 
    (0,0) node[circle, draw, fill=black!50,
                        inner sep=0pt, minimum width=3pt, label=below:{$v_1$}] {};
    \draw [fill=blue!25] 
    (1,0) node[circle, draw, fill=black!50,
                         inner sep=0pt, minimum width=3pt, label=below:{$-v_0$}]  {};
\end{tikzpicture}}     \\
\partial_1([v_0,v_1]) &= [v_1] - [v_0]. 
           \end{aligned}
            \end{equation*}        
For a 2-simplex, we compute
                \begin{equation*}
                        \begin{aligned}
                                \partial_2\left(
                                                               \raisebox{-0.8cm}{
\begin{tikzpicture}[bullet/.style={circle, fill=black!50, inner
sep=0pt, minimum width=3pt, label={[font=\bfseries]#1}},>=latex]
    \draw [fill=blue!25] 
    (210:.5) node[bullet={below left:$v_0$}] (a) {}
    -- (90:.5) node[bullet={above:$v_1$}] (b){}
        -- (-30:.5) node[bullet={below right:$v_2$}] (c) {}
            -- cycle;
    \draw[-{Latex[bend]}] (240:0.15) arc(240:-60:0.15); 
\end{tikzpicture}}
 \right) &= \left(
                                                                \raisebox{-0.8cm}{
\begin{tikzpicture}[bullet/.style={circle, fill=black!50, inner
sep=0pt, minimum width=3pt, label={[font=\bfseries]#1}},>=latex]
    \draw [black,line width=.5pt,->-/.list={1/2}] 
     (90:.5) node[bullet={above:$v_1$}] (b){}
        -- (-30:.5) node[bullet={below right:$v_2$}] (c) {};
\end{tikzpicture}}
-
                                                               \raisebox{-0.8cm}{
\begin{tikzpicture}[bullet/.style={circle, fill=black!50, inner
sep=0pt, minimum width=3pt, label={[font=\bfseries]#1}},>=latex]
    \draw [black,line width=.5pt,->-/.list={1/2}] 
    (210:.5) node[bullet={below left:$v_0$}] (a) {}
        -- (-30:.5) node[bullet={below right:$v_2$}] (c) {};
\end{tikzpicture}}
+
                                                               \raisebox{-0.8cm}{
\begin{tikzpicture}[bullet/.style={circle, fill=black!50, inner
sep=0pt, minimum width=3pt, label={[font=\bfseries]#1}},>=latex]
    \draw [black,line width=.5pt,->-/.list={1/2}] 
    (210:.5) node[bullet={below left:$v_0$}] (a) {}
    -- (90:.5) node[bullet={above:$v_1$}] (b){};
\end{tikzpicture}}
      \right)\\
      \partial_2([v_0,v_1,v_2]) &= [v_1,v_2]-[v_0,v_2]+[v_0,v_1] = [v_1,v_2] +[v_2, v_0] + [v_0,v_1],
                        \end{aligned}
            \end{equation*}              
            since $[v_0,v_1] = -[v_1,v_0].$
\end{ex}

An $\sdim$-chain~$c$ with empty boundary $\partial c = 0$ is called an \emph{$\sdim$-cycle}. 
The $\sdim$-cycles form a subgroup $Z_\sdim:=\ker \partial_\sdim$ of the {$\sdim$th chain group}.
An \emph{$\sdim$-boundary} is an $n$-chain that is the boundary of an $(\sdim+1)$-chain $c = \partial d$ with $d \in C_{\sdim+1}(\Si,\F)$; 
the \emph{$n$-boundaries} $B_\sdim:= \ima \partial_{\sdim+1}$ are a subgroup of the $\sdim$th chain group. 
{The chain groups and linear maps $\{C_\sdim(\Si, \F),\partial_\sdim\}:=\cdots  \overset{\partial_{3}}{\longrightarrow} C_{2}(\Si, \F) \overset{\partial_{2}}{\longrightarrow} C_1(\Si, \F) \overset{\partial_{1}}{\longrightarrow} C_{0}(\Si, \F)\overset{\partial_{0}=0}{\longrightarrow} 0$ form a \emph{chain complex} if the image of each map is included in the kernel of the next.}

\begin{ex}\label{ex:chaincomplex}(\cref{example} continued) {Consider the simplicial complex from Example~\ref{example}.
The collection of vertices $\Si_0=[1,2,3,4,5]$ is a basis for~$C_0(\Si,\F)$ so the elements take the form of linear combinations of the vertices with coefficients in the field. 
We follow convention to write the chain complex with the field we are working over: $\F_{\sdim}^{\alpha_\sdim}$ where~$\alpha_\sdim$ is the number of $\sdim$-simplicies and~$\sdim$ indexes the $\sdim$th chain group.
Since~$\Si$ is a simplicial complex of dimension~2, then for all $p>\sdim$, we have $\Si_p$ empty and $\F_p=0$. In Example~\ref{example}, we can write the chain complex:
\[ 0 \overset{\partial_{3}}{\longrightarrow} \F_{2}^{1}(\Si) \overset{\partial_{2}}{\longrightarrow} \F_1^{7}(\Si)  \overset{\partial_{1}}{\longrightarrow} \F_0^5 (\Si) \overset{\partial_{0}=0}{\longrightarrow} 0.\]}
\end{ex}

It can be proven that the composition of any two consecutive boundary maps is the zero map, that is, $\partial_\sdim \circ \partial_{\sdim+1} = 0$ for all~$\sdim$, which means that $B_\sdim=\ima\partial_{\sdim+1} \subseteq\ker\partial_\sdim = Z_\sdim$. This result is known as the \emph{Fundamental Lemma of Homology}, 
and shows that the chain groups~$C_\sdim(\Si, \F)$ together with the boundary maps $\partial_\sdim$ form a \emph{chain complex}~\cite{Hatcher2002}.

\begin{ex}
It is easy to see in Example~\ref{ex:boundary} that the composition of two boundary maps yields zero, i.e.,
  \begin{equation*}
  \begin{split}
     \partial_1( \partial_2([v_0,v_1,v_2])) &= \partial_1([v_1,v_2] +[v_2, v_0] + [v_0,v_1]) \\ &= [v_2] - [v_1] + [v_0] - [v_2] + [v_1] - [v_0] = 0.
     \end{split}
\end{equation*}
\end{ex}
Simplicial homology now quantifies the $\sdim$-cycles that are not boundaries.
Intuitively speaking, these special $\sdim$-cycles represent $n$-dimensional ``holes.''
For instance, in~\Cref{example}, the $1$-chain $[3,4] + [4,5] - [3,5]$ is clearly a cycle, but it is not the boundary of any $2$-dimensional chain.

\begin{defn}[$\sdim$th simplicial homology group]\label{def:homology}
Let~$\Si$ be a simplicial complex. For $\sdim\geq 0$, the $\sdim$th \emph{homology group} is
\begin{align*}
H_\sdim(\Si,\F) := Z_\sdim/B_\sdim = {\ker \partial_\sdim}/{\ima \partial_{\sdim+1}}.
\end{align*}
\end{defn}

The dimensions of the homology groups, which are the ranks of the corresponding vector spaces, describe different topological features. 
Specifically, the elements of the homology groups~$H_0$,~$H_1$ and~$H_2$ describe \emph{connected components}, \emph{loops}, and \emph{voids} of~$\Si$, respectively. 
Two elements in~$H_\sdim$ are different if they differ by more than a boundary. Then these two elements belong to different homology classes and these different classes represent two distinct $\sdim$-dimensional holes. For example, one-dimensional loops in the same homology class all surround the same one-dimensional hole.
The number of $\sdim$-dimensional holes is called the $\sdim$th \emph{Betti number} 
\begin{align*}
\beta_\sdim := \dim H_\sdim(\Si,\F) = \dim \ker \partial_\sdim - \dim \ima \partial_{\sdim+1}.
\end{align*}
The first three Betti numbers~$\beta_0$, $\beta_1$, and~$\beta_2$ represent, respectively, the number of connected components, the number of $1$-dimensional holes, and the number of $2$-dimensional holes (i.e., voids) in a simplicial complex. 
The Betti numbers also relate to the Euler characteristic through the Euler-Poincar\'e theorem \cite[Theorem 2.44]{Hatcher2002}.

\begin{rem}
Simplicial homology can also be defined with coefficients in a ring instead of a field; indeed, simplicial homology is defined with integer coefficients in Refs.~\cite{Hatcher2002,munkres2018elements}. In this case, the definitions above generalize to modules and homomorphisms.
Changing the coefficient field can affect the Betti numbers. For example, the homology of the Klein bottle with coefficients in $\Z/2\Z$ has $\beta_1=2$ and $\beta_2=1$ but the homology of the Klein bottle for all other coefficients in $\Z, \R,$ or $\Z/p\Z$ where p is prime, is $\beta_1=1$ and $\beta_2=0$.
\end{rem}

To calculate homology in practice, there are computer packages available such as: \texttt{SimplicialComplexes.m2}
package in Macaulay2~\cite{M2} to compute homology with different coefficients; the simcomp toolbox in GAP~\cite{effenberger2011simpcomp}; and the topology application \texttt{`topaz'} in polymake~\cite{gawrilow2000polymake}. Persistent homology computations use other packages described in the next section.

\subsection{Persistent homology}
\label{sec:PersHom}
Persistent homology (PH) is one of the prominent tools in topological data analysis. PH enables the quantification of meaningful topological and geometric features in data by studying homology across multiple scales of data. 

\begin{defn}[See~\cite{carlsson2009,edelsbrunner2008,Ghrist2008}]
A \emph{filtration} of a simplicial complex~$\Si$ is a sequence of nested simplicial complexes,
\begin{equation*}
	\emptyset = \Si_0 \subseteq \Si_1 \subseteq \Si_2 \subseteq \dotsb \subseteq \Si_{\mathrm{end}} = \Si,
\end{equation*}
starting with the empty complex and ending with the entire simplicial complex~$\Si$. 
\end{defn}

For a given filtration, for example, the \v{C}ech complex~$\Si_\eps$ outlined in \Cref{ssec:topology_preview}, one can track when homology groups  appear and disappear throughout the filtration. 
For instance, we can keep track of the number of connected components while varying the parameter:
If the number of components persists over a large range of the scale parameter, this may be indicative of a clustered dataset. 

One of the key properties that allows the PH pipeline to take in a filtration and output a topological summary is functoriality. Homology is \emph{functorial}: Any map between simplicial complexes $f_{i,j}: \Si_i \rightarrow \Si_j$ induces a map between their $\sdim$-chains $\tilde{f}^\sdim_{i,j}: C_\sdim(\Si_i,\F)~\rightarrow~C_\sdim(\Si_j,\F)$, which then induces a map between their homology groups $f^\sdim_{i,j}: H_\sdim(\Si_i,\F) \rightarrow H_\sdim(\Si_j,\F)$. Since the simplicial complexes in a filtration are related by inclusion, there exists maps between the homology groups of every simplicial complex in a filtration. This enables one to track when topological features appear and disappear throughout the filtration: A topological feature of dimension~$n$ in~$H_\sdim(\Si_{b},\F)$ is \emph{born} in~$H_\sdim(\Si_{b},\F)$, if it is not in the image of~$f^\sdim_{b-1,b}$ and a feature from $H_\sdim(\Si_i)$ \emph{dies} in~$H_\sdim(\Si_d,\F)$, where $i<d$, if~$d$ is the smallest index such that the feature is mapped to zero by $f^\sdim_{i,d}$. 
Thus, such a topological feature can be associated to a half-open interval~$[b,d)$ and the lifetime of a topological feature, i.e., its length $d-b$, is called its \emph{persistence}. 
A multiset of such intervals is called a topological \textit{barcode}. 
Thus, for an appropriate choice of basis~\cite{Zomorodian2005} of the homology groups~$H_\sdim$, a barcode summarizes the information carried by the homology groups and the maps~$f^\sdim_{i,j}$~\cite{carlsson2009,Ghrist2008}, its \emph{persistent homology}.

Persistent homology benefits from theoretical underpinnings that prove a barcode is robust to small perturbations to input data~\cite{cohen2007stability}. 
Persistent homology is computable, with many software implementations available, and we refer to the following tutorials and ``user guides'' to more information~\cite{otter2017roadmap,ray2017survey,munch2017user,chazal2021introduction,patania2017topological}. 

\begin{rem}
It is also possible to define simplicial cohomology and compute barcodes for the cohomology groups for a given filtration of a simplicial complex~$\Si$. In general, cohomology is more powerful than homology. But for coefficients in a field~$\F$, cohomology groups are dual to homology groups, and more generally, they are related by the Universal Coefficient Theorem. 
In the setting we consider here, homology and cohomology yield identical barcodes~\cite[Prop.~2.3]{de2011dualities}, which is why one usually does not distinguish persistent homology and persistent cohomology.
\end{rem}

Mathematical theory extending persistent homology is still an active area of research. 
For example, when data cannot be encoded as a filtration (i.e., a nested sequence of spaces), or if the dataset is better analysed with two parameters (i.e., a bi-filtration), then zig-zag persistence~\cite{carlsson2009zigzag,carlsson2010zigzag} or multiparameter persistent homology~\cite{carlsson2009theory,carlsson2009computing}, respectively, may be more suitable. 
The computation of these and other homological objects are actively developed and implemented to be accessible for applications~\cite{henselman2016matroid,carlsson2019persistent,miller2017data,scaramuccia2020computing,harrington2019stratifying,kerber2021fast,lesnick2015interactive}. 
An active area of research for studying higher-order analogues of directed networks is the mathematical field of directed algebraic topology, which aims to capture periodicity in data~\cite{fajstrup2016directed}.
In recent years, persistent homology of {certain} directed complexes~\cite{lutgehetmann2020computing}, persistent path homology~\cite{chowdhury2018persistent,dey2020efficient}, and weighted path homology have also been proposed~\cite{lin2019weighted}.

Ultimately, to analyze data, we need to compare outputs of persistence computations.
The most common approach to achieve this is to transform the barcodes into a vector-format, which can take various forms, such as persistence images or persistence landscapes, or can be based on kernel methods, which can then be interpreted with tools from statistics and machine learning~\cite{adams2017persistence,kwitt2015statistical,chazal2015subsampling,bubenik2015statistical}. 
Developing such statistical tools for topology, including hypothesis testing, bootstrapping, and defining shape statistics, is another important, related research area~\cite{robinson2017hypothesis,fasy2014introduction,wasserman2018topological,Fasy2014,reani2021cycle}.

\subsection{Applications of Persistent Homology}
Persistent homology has found widespread application to real-world complex systems%
\footnote{A collection of TDA applications is also available at \url{https://www.zotero.org/groups/2425412/tda-applications}}. 
For example, persistent homology has also been used to describe 3D structures in material science~\cite{saadatfar2017pore,hiraoka2016hierarchical} as well as to analyze sensor networks (see two accessible introductions~\cite{de2007homological,robinson2019hunting}.
Different applications have different types of input data, such as point clouds in a high dimensional space, network data (nodes and relations), 
images (grayscale pixel data in a square grid), or data as a given hypergraph. 

When data is given as a point cloud, one needs to determine what is the appropriate filtration parameter and corresponding complex to build~\cite{ghrist2014elementary,carlsson2009,edelsbrunner2010computational,edelsbrunner2008}. 
For persistent homology, the chosen filtration is highly dependent on the application (see, for example, \cite{stolz2014computational,stolz2019global}), and the filtration may limit the user to a particular software. While the homology or persistent homology of hypergraph data is possible~\cite{bressan2016embedded}, ~\cite{ren2020persistent}, this requires a suitable filtration. The applications discussed in this section focus on simplicial complexes or its 1-skeleton (a graph), which are both types of hypergraphs.

Using topological methods to study problems in neuroscience was first proposed in the 1960s~\cite{zeeman1962topology}. The availability of neuroscience data as well as advances in mathematical theory and computation have enabled this field to flourish. Computational topology and higher-order networks have proven successful for analyzing the full spectrum of brain data ranging from functional networks~\cite{sizemore2019importance}, to morphology of branching neurons~\cite{kanari2018topological}, to structural (synaptic connectivity)~\cite{reimann2017cliques}, to place cells~\cite{giusti2015clique}, to the \textit{C. elegans} connectome~\cite{helm2020growing}, to imaging of brain disease~\cite{crawford2016topological,bruningk2020image}. Rather than an exhaustive list of research in topological neuroscience, we refer to the reader to a few recent survey articles~\cite{curto2017can,hess2020topological,blevins2020topology}.

Higher-order networks have also proven useful in genomics and evolutionary biology~\cite{rabadan2019topological,chan2013topology}, structural biology~\cite{xia2014persistent}, as well as for the analysis of structures such as vascular networks~\cite{bendich2016persistent,stolz2020multiscale,byrne2019topology}. 
References overviewing the potential of topological techniques (eg to biology) include~\cite{blevins2020topology,amezquita2020shape,rabadan2019topological,stolz2019global}.
Recent studies suggest that higher-order network structures and computational topology can be helpful for analyzing complicated mathematical models of biological systems~\cite{topaz2015topological,kim2020spatiotemporal,nardini2021topological,mcguirl2020topological}.
Using higher-order networks to analyze diseases such as cancer~\cite{aukerman2020persistent,nicolau2011topology,stolz2020multiscale,gonzalez2020prediction} offers possibilities for combining data and mathematical models~\cite{stolz2017persistent,vipond2021}. 
While the structure of some chemical reaction models can be distinguished using persistent homology~\cite{vittadello2020model}, others are better encoded as a hypergraph and analyzed with discrete Ricci curvature~\cite{eidi2020edge}. 

More generally, many biological processes, including those in neuroscience, can be modeled as dynamical systems, which we can also try to understand using topological tools.
Indeed studying relations between topology and dynamics has deep roots, going back to Morse-Smale dynamical systems. 
Tools for analyzing time series data and dynamical systems using topological methods have been developed by Harer, Perea, Robins, Mischaikow, Mukherjee among others (see~\cite{perea2019topological,robins1998computing,harer2015inferring,kaczynski2006computational,mischaikow2002conley} and references within). 
Memoli, Munch, and colleagues have been extending persistent homology theory and methods to analyse time varying systems, ranging from collective behavior in the form of dynamic point cloud, dynamic graphs, and Hopf bifurcation detection~\cite{munch2013applications,kim2017stable,kim2020analysis,tymochko2020using}.
Growing graphs have also been analysed with node-filtered order complexes~\cite{blevins2020reorderability}. 
Time varying data has been analyzed with topological summaries, such as vineyards, crocker plots, and multiparameter rank functions~\cite{xian2020capturing}, whereas temporal networks have been analysed with persistent path homology~\cite{chowdhury2020path}. 
Contagion dynamics on different network structures (e.g., ring lattice and tori) as well as simplicial complexes have been analysed with persistent homology~\cite{taylor2015topological,Iacopini2018}. 
Dynamics also occur on a more general hypergraphs, which motivates approaches beyond persistent homology~\cite{carletti2020dynamical}.

\subsection{Hodge-Laplacians and Hodge-theory}
\label{sec:Hodge}
Persistent homology is one of the work horses of topological data analysis.
As discussed, persistent homology reveals homological information in a given dataset by analyzing the sequence of boundary maps~$\partial_n$ when scanning through a filtration.
An interesting alternative way to use the boundary maps $\partial_n$ to extract information about a simplicial complex is by constructing \emph{Hodge-Laplacians}~\cite{lim2020hodge,Hatcher2002,eckmann1944harmonische,grady2010discrete,horak2013spectra}
\begin{equation}\label{eq:hodge_laplacian}
    \bm L_n =  \bm W_{n+1}^{} \bm B_n^\top \bm W_n^{-1} \bm B_n^{} + \bm B_{n+1}^{}\bm W_{n+2}^{}\bm B_{n+1}^\top\bm W_{n+1}^{-1},
\end{equation}
where $\bm B_n$ are the matrix representation of the boundary operators~$\partial_n$, and~$\bm W_n$ are (diagonal) weight matrices.

\begin{ex}
    Consider the simplicial complex shown below. 
    \begin{center}
    \begin{tikzpicture}
        \small
        \tikzstyle{point}=[circle,thick,draw=black,fill=white,inner sep=2pt,minimum width=10pt,minimum height=10pt]
        \tikzstyle{smallpoint}=[circle,thick,draw=black,fill=black,inner sep=0pt,minimum width=5pt,minimum height=5pt]
        \begin{scope}[>=latex, thick, decoration={ markings, mark=at position 0.5 with {\arrow{>}}},font=\tiny,every node/.style={sloped}]
            \node (1) at (0,0) [point] {\normalsize $1$};
            \node (2) at (1.5,0) [point] {\normalsize $2$};
            \node (3) at (0.75,-1) [point] {\normalsize $3$};
            \node (4) at (2.25,-1) [point] {\normalsize $4$};

            \node (5) at (3.75,-1) [point] {\normalsize $5$};
            \node (6) at (3.0,0)    [point] {\normalsize $6$};

            \draw[postaction={decorate}] (1) -- node[above]{$[1,2]$} (2);
            \draw[postaction={decorate}] (1) -- node[below]{$[1,3]$} (3);

            \draw[postaction={decorate}] (2) -- node[below]{$[2,3]$} (3);
            \draw[postaction={decorate}] (2) -- node[above]{$[2,4]$} (4);
            \draw[postaction={decorate}] (2) -- node[above]{$[2,6]$} (6);

            \draw[postaction={decorate}] (3) -- node[below]{$[3,4]$} (4);

            \draw[postaction={decorate}] (4) -- node[below]{$[4,5]$} (5);

            \draw[postaction={decorate}] (5) -- node[above]{$[5,6]$} (6);

            \draw[thin,->] ([yshift=-0.2cm] barycentric cs:1=1,2=1,3=1) arc (280:0:0.2cm);
            \begin{pgfonlayer}{background}
                \fill[opacity=0.3,gray] (1.center) -- (2.center) -- (3.center) -- cycle;
            \end{pgfonlayer} 
        \end{scope} 
    \end{tikzpicture}
    \end{center}

    In this case, the boundary maps $\bm{B}_1$ (rows indexed by nodes, columns indexed by edges) and $\bm{B}_2$ (rows indexed by edges, columns indexed by $2$-simplices) are:
    \begin{equation*}
        \small
        \arraycolsep=1.4pt\def\arraystretch{1}
        \bm{B}_1 = 
        \begin{array}{c|cccccccccc}
            \;& [1,2] & [1,3] & [2,3] & [2,4] & [2,6] & [3,4] & [4,5] & [5,6]\\
                \hline 
             1  &   -1 & -1 & 0 & 0 & 0 & 0 & 0 & 0 \\
             2  &   1 & 0 & -1 & -1 & -1 & 0 & 0 &  0 \\
             3  &   0 & 1 & 1 & 0 & 0 & -1 & 0 &  0 \\
             4  &   0 & 0 & 0 & 1 & 0 & 1 & -1 & 0\\
             5  &   0 & 0 & 0 & 0 & 0 & 0 & 1 &  -1 \\
             6  &   0 & 0 & 0 & 0 & 1 & 0 & 0 &  1 \\
            \end{array}\;\;\;
        \bm{B}_2 = 
        \begin{array}{c|cc}
            \; & [1,2,3] \\
            \hline
            [1, 2] & 1 \\
            \left [1, 3\right] & -1 \\
            \left [2,3\right ] & 1 \\
            \left [2,4\right ] & 0 \\
            \left [2,6\right ] & 0 \\
            \left [3,4\right ] & 0 \\
            \left [4,5\right ] & 0 \\
            \left [5,6\right ] & 0 \\
        \end{array}
    \end{equation*}
\end{ex}

Note that for graphs we can recover the standard graph Laplacian~\cite{chung1997spectral} as $\bm L_0 = \bm B_1 \bm B_1^\top$, where $\bm B_1$ is simply the signed node-edge incidence matrix known from graph theory, $\bm B_0 :=\bm 0$, and $\bm W_2=\bm W_1=\bm I$.
As a second example, the random walk Laplacian~\cite{masuda2017random} can be also be recovered via $\bm L_\text{rw} = \bm B_1\bm B_1^\top\bm W_1^{-1}$, where $\bm W_2=\bm I$ and $\bm W_1$ is set to the diagonal matrix of node degrees.
{We remark that the weight matrices $\bm W_i$ can often be estimated from data as well, as has been done, e.g., as done in~\cite{chen2021helmholtzian} the context of estimating the $\bm L_1$ Hodge Laplacian (Helmholtzian).
We finally remark that can define a symmetric variant via the similarity transform $\bm W_{n+1}^{-1/2}\bm L_n\bm W_{n+1}^{1/2}$~\cite{horak2013spectra,Schaub2020,grady2010discrete}, from which it is easy to see that the Hodge Laplacian~$\bm L_n$ is positive semi-definite for any $n$.}

Based on the Hodge Laplacians, as defined in Equation~\cref{eq:hodge_laplacian}, we can compute most information about the homology of a (simplicial) complex:
For instance, we may compute Betti numbers~\cite{friedkin1991theoretical} by assessing the null space of the Hodge-Laplacian.
To understand why this is the case, it is insightful to introduce the Hodge-decomposition~\cite{lim2020hodge,grady2010discrete,Schaub2021}.
For simplicity we concentrate here on the unweighted case where $\bm W_n=\bm I$ for all~$n$, though weighted Hodge-decompositions are also available~\cite{Schaub2020}.
In this case the Hodge decomposition states that the space of $n$-chains (cochains) can be decomposed into three orthogonal subspaces, that is,
\begin{equation}\label{eq:Hodge_decomposition}
    C_n(\Si,\F)  \cong \ima(\bm B_n^\top) \oplus \ima(\bm B_{n+1}) \oplus \ker(\bm L_n),
\end{equation}
where~$\oplus$ represents the direct sum of orthogonal subspaces.
The three subspaces of the Hodge-decomposition are generally spanned by certain subsets of eigenvectors of~$\bm L_n$~\cite{barbarossa2020topological,Schaub2020,Schaub2021}.
Importantly, we have~\cite{lim2020hodge}
\begin{equation}
    \ker(\bm L_n) \cong \ker(\bm B_n)\cap \ker(\bm B_{n+1}^\top) \cong \ker(\bm B_n) / \ima(\bm B_{n+1})
\end{equation}
and hence the kernel of the $n$th Hodge Laplacian $\ker(\bm L_n)$ is isomorphic to the $n$th homology vector space~$H_n$ as given in~\Cref{def:homology}.

Moreover, Hodge Laplacians, enable the rigorous definition of local dynamical processes, such as diffusion and consensus dynamics in the domain of edges (node-pairs) and higher-order entities~\cite{muhammad2006control,lim2020hodge,horak2013spectra,Schaub2020,parzanchevski2017simplicial,parzanchevski2016isoperimetric}. 
These ideas are also closely connected to signal processing on a (simplicial) complex~\cite{grady2010discrete,Barbarossa2020,barbarossa2020topological,Barbarossa2018,barbarossa2016introduction,Schaub2021,yang2021finite,yang2022simplicial,schaub2022signal,roddenberry2022signal}, where one can define operations such as convolutions based on such Hodge-Laplacian operators, or even define analogues of graph-neural networks for simplicial complexes~\cite{roddenberry2019hodgenet,bodnar2021weisfeiler, ebli2020simplicial,glaze2021principled,bunch2020simplicial,billings2019simplex2vec,hacker2020k}.  
Based on related ideas, one can moreover construct decentralised protocols to solve coverage problems in sensor networks~\cite{tahbaz2010distributed}, rank nodes within a network~\cite{Jiang2011}, or decompose games~\cite{candogan2011flows}.

\section{Understanding relational data with higher-order network models}\label{sec:Statistics}
{In this section, we describe how (higher-order) networks are used to study relational data.}
Before embarking on a more extended discussion, we emphasize in~\Cref{ssec:rel_data_vs_relations_from_data}  differences with the perspective on higher-order models discussed so far, before giving a more detailed discussion on what we mean mathematically with higher-order data in~\Cref{ssec:higher_rel_data}.
We then discuss various probabilistic models of such data based on simplicial complexes and hypergraphs in \Cref{ssec:prob_models}, and point out several application scenarios in~\Cref{ssec:stat_examples}.

\subsection{Relational data \emph{vs} relations created from data}\label{ssec:rel_data_vs_relations_from_data}
Modeling relations among data is of course a theme of this review, discussed with respect to topology and geometry in~\Cref{sec:Structure} and dynamics in the following~\Cref{sec:Dynamics}.
However, when we turn our focus towards relational data itself, there are differences as to what we aim to understand as well as what mathematical tools are typically used.
Accordingly, the role (higher-order) networks play in the analysis is somewhat different.

\Cref{sec:Structure,sec:Dynamics} are effectively concerned with understanding how data (e.g., time-series) associated to \emph{nodes} in a network are organized topologically, geometrically, or dynamically, and how we can use relations between those nodes to understand that data better.
Observe that at the core of the ``geometry and topology'' scenario described in~\Cref{ssec:topology_preview} is the idea of \emph{constructing} appropriate graphs and higher-order objects to understand  certain topological and geometric questions about data (often point cloud data).
For instance, a graph, in which each data point becomes a node, is constructed according to a nearest neighbor rule from measured point-cloud data.
Then the eigenvectors of the resulting graph Laplacian are used as a new coordinate system to describe the data.
Similarly, the essential question when considering dynamical networks (\Cref{ssec:dynamics_preview}) is how we can comprehend the (global) dynamics of the system by analyzing the underlying higher-order network.

In contrast, when modeling relational data we are typically not interested in understanding data associated to nodes, but the data we measure and aim to model comes (directly) in the form of \emph{relations}, such as friendship links or collaborations.
For instance, we may want to model a collaboration network by specifying how likely it is for each collaboration to occur.
Stated differently, the data to model is the set of (hyper)edges or faces of a (higher-order) network or complex.
While it is also possible to try to \emph{infer} such relations from nodal observations only, e.g., by measuring correlations between node variables, in the following, we will be primarily concerned with scenarios in which the relations of interest can directly be measured.
This encapsulates a broad set of data, including co-occurence relationships, collaborations, affiliations, or flows of mass, energy, money, etc.

\newcommand{\maxdata}{m}
\newcommand{\maxsam}{D}

\newcommand{\maxedg}{M}

\subsection{Describing higher-order relational data}\label{ssec:higher_rel_data}
Let us assume we are given a finite set of data $\mathcal D = \sset{X_i}$
consisting of some measured relations~$X_i$ between a finite number of entities $v_1, \dotsc, v_\maxdim$, which we will represent as vertices.
For concreteness, let us use an example in which~$\mathcal D$ is the set of all collaboration relationships within a large number of scientific papers.
In this example, the vertices~$v_i$ will correspond to authors and the relations~$X_i$ will be finite collections of co-authors, where we have one measured relation for each paper in our data~$\mathcal D$.

When modeling such data as a dyadic network, we typically preprocess the data by first splitting all sets~$X_i$ into their subsets of size two, if required (cf.~\cref{fig:schematic_higher_order_rel_data}). 
In contrast, higher-order models avoid such preprocessing and explicitly account for the non-dyadic relationships.
Even though we thus keep all measured relationships in the data intact, there some modeling choices to be made, as to how we represent the higher-order relations.
The most prominent choices here are to either use a hypergraph model, or think of the data as defining an (abstract) simplicial complex.
Depending on the data and the question under consideration, one or the other representation may be more useful than the other.

Specifically, in the simplicial complex model we assume that for every relation~$X_i$ all subsets $Y_i \subset X_i$ are part of the complex as well.
This may or may not be a reasonable modeling assumption, depending on the context.
In our author-collaboration example, the subset inclusion assumption inherent to a simplicial complex can be appropriate: If authors~$A$,~$B$ and~$C$ collaborated on a paper~$X_i$, this implies that~$A$ and~$B$ collaborated in the context of~$X_i$, too.
However, if the relation to be modelled is simply a joint paper, then the simplicial abstraction may be misleading: a joint three author paper does not imply that there are (at least) three papers between all possible pairs of authors.
There are also other contexts in which a simplicial complex may not be applicable as model.
For instance, assume we want to model a set of chemical reactions~$X_i$, comprising different chemical species, as a higher-order network.
The fact that that species~$A, B$ and~$C$ are part of a joint reaction to form some new compound~$D$, does not imply that~$A$ and~$B$ will react with each other in the absence of~$C$.
For instance, $C$~could be an enzyme that mediates the reaction, and in its absence no reaction between~$A$ and~$B$ takes place.

Notably, the choice of the modeling abstraction of the data, has strong consequences for the tools available for downstream analysis.
If the system can be aptly represented by a simplicial complex, then we have a large array of tools from applied topology and topological data analysis available with which we can investigate various characteristics of the data, such as question about homology (e.g., are there gaps in the space of author-collaborations?).
In contrast, comparably less tools are available for hypergraphs, even though they provide a more flexible abstraction of the data, in principle.
We refer to the recent overview~\cite{torres2020representations} for a more in-depth discussion about relevant (higher-order) representation for modeling data.

\subsection{Probabilistic models for higher-order networks}\label{ssec:prob_models}
Unless our goal is only to perform a descriptive analysis, once we have chosen an appropriate framework to represent the higher-order data, we typically need to define an appropriate probabilistic, {or statistical}, model for such data for further statistical analysis.
In the following  we discuss some directions in the study of probabilistic models for higher-order networks.
Rather than provide a complete discussion, our goal here is to highlight some models that have prompted further mathematical inquiries.
For a more exhaustive overview, one can refer to~\cite{Battiston2020}.

{
As a starting point, and to set-up the language of this section, in Example \ref{ex:model-examples}, we explore the Erd\H{o}s-R\'{e}nyi random graph model and introduce generalizations of this model to higher-order networks that we will describe in more detail later in the section.  To begin, we formally define a \emph{statistical model}. 
In this section, when we use the term \emph{model} or \emph{probabilistic model}, we are often referring formally to statistical models and have this definition in mind.
}

\begin{defn} \label{def:statistical-model} 
{
A \emph{statistical model}~$\mathcal M$ is a collection of probability distributions on the same sample space.  
    In the discrete setting, a statistical model~$\mathcal M$ on a sample space of size~$\maxsam$ is a subset of the $\maxsam-1$ dimensional probability simplex $\Delta_{\maxsam-1}$, i.e.,
\[\mathcal M \subseteq \Delta_{\maxsam-1}: = \set{q \in \mathbb R^{\maxsam} }{\sum_{i=1}^\maxsam q_i =1, \ q_i \geq 0 \text{ for all }i }.\]
}
\end{defn}

\begin{figure}[tb!]
    \centering
    \includegraphics[width=3in]{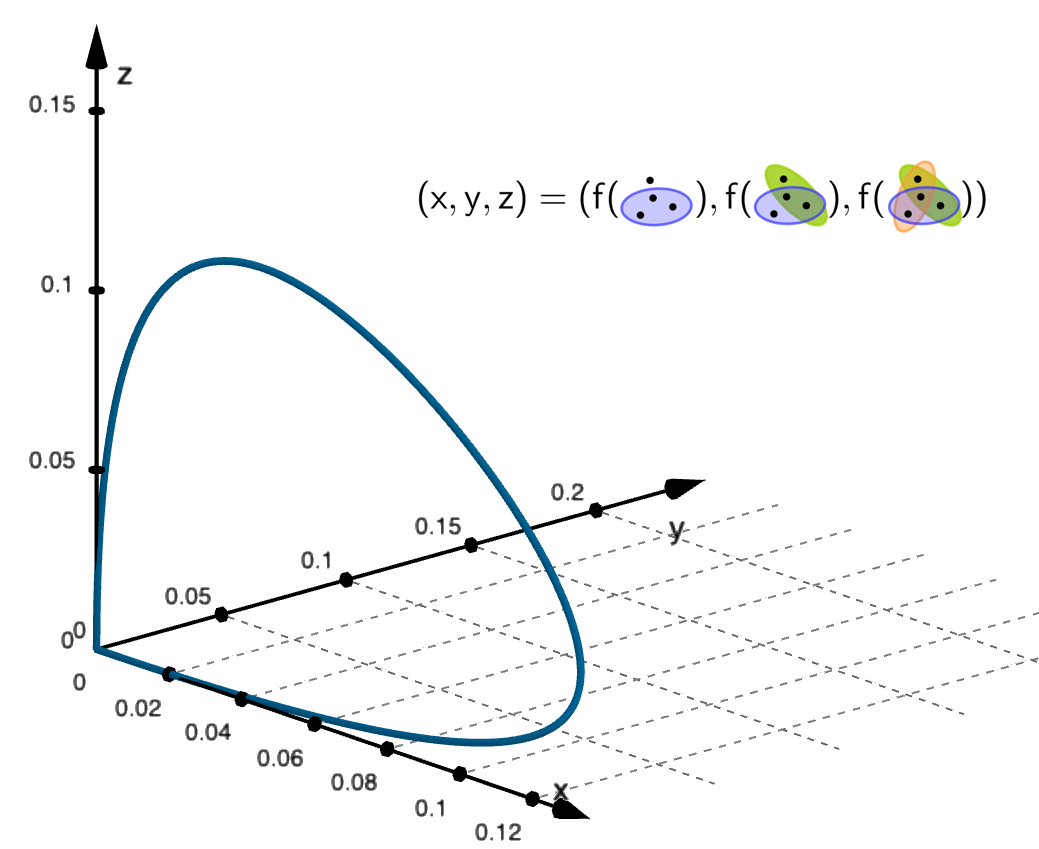}
    \caption{ A projection of the one-dimensional random hypergraph model~$H^3(4, p)$ onto three of its coordinates.  Each point in the model is a discrete distribution (probability mass function) ~$f$ over the set of all~16 possible 3-uniform hypergraphs on 4~vertices.  In this projection, we follow three of the coordinates as the parameter $p$ varies from $0$ to $1$.} %
        \label{fig:schematic_projection}
\end{figure}

\begin{ex} \label{ex:model-examples}
{
Let $\mathcal G_\maxdim$ be the space of all simple graphs on~$\maxdim$ vertices. 
    The Erd\H{o}s-R\'enyi random graph model $\mathcal M_\maxdim = G(\maxdim,p)$, where each edge on~$\maxdim$ vertices is chosen with probability~$p$, is a statistical model as in Definition~\ref{def:statistical-model} parameterized by a single parameter~$p$.  
    }

{
Let $\maxdim=3$. In this case, $\mathcal M_3 \subseteq \Delta_{7}$ where $7 = 2^{ {\maxdim \choose 2}}-1$. Given $p$, the probability distribution of observing each of the 8 graphs in $\mathcal G_\maxdim$ is given by the vector 
$\ $\\
\begin{center}
    \includegraphics[width=5in]{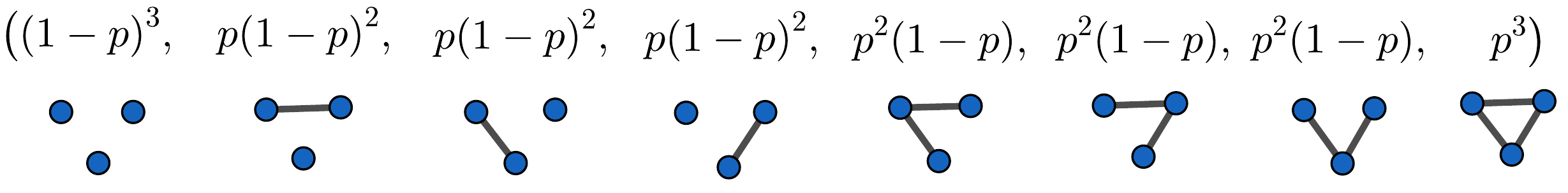}.
\end{center}
}

{
Generalizations of the random graph model~$G(\maxdim,p)$ for simplicial complexes and hypergraphs  are discussed in \Cref{subsec:randomsimplicialcomplexes,subsec:prob-models-hypergraphs}, respectively.  
For hypergraphs, one generalization of~$G(\maxdim,p)$ is the $H^d(\maxdim,p)$ model, a statistical model on the set of all $d$-uniform hypergraphs on~$\maxdim$ vertices where each $d$-uniform hyperedge is selected with probability~$p$.  When $d=3$ and $\maxdim=4$ the sample space corresponding to the model has size~$2^{{4 \choose 3}}=16$.  A projection of the one-dimensional model~$H^3(4, p)$ onto three of the 16 coordinates is displayed in Figure~\ref{fig:schematic_projection}.
}
\end{ex}

\subsubsection{Probabilistic models for simplicial complexes and cell complexes} \label{subsec:randomsimplicialcomplexes}
The study of random simplicial and cell-complexes is still a relatively nascent area of research.
Two main lines of inquiry here can be distinguished here.
First, there are random models for complexes that may be seen as generalizations of random graph models such as the famous Erd\H{o}s-R\'enyi (ER) random graph model {discussed above}.
We refer to~\cite{Kahle2014,Costa2012,Kahle2013} for early surveys over some of these generalizations and results.
Second, there are a number of models for random geometric complexes.
In contrast to the first models, which may also be seen as models for random abstract (simplicial) complexes, the construction of these geometric models is often much closer to the setup considered in topological data analysis (see~\Cref{sec:Structure}) and thus of particular relevance as null models for this class of applications.

\paragraph{Random abstract simplicial complexes} One of the earliest models in this direction was proposed by Linial and Meshulam~\cite{Linial2006}.
Their model starts with a complete graph (a complete 1-skeleton) and then adds 2-simplices independently at random with a probability~$p$. 
They show that just like there is threshold phenomenon with respect to the zeroth homology group (i.e., connected components) of any sampled graph in the ER model {(i.e., the emerge of a giant connected component and an almost surely connected graph, once certain connection probabilities are exceeded)}, there exists a corresponding threshold phenomenon with respect to the first homology group in their random 2-complex model.
These results got refined and extended later~\cite{Babson2011,Hoffman2012,Luczak2018}, for instance~\cite{Meshulam2009} presents analogous results for random $\sdim$-dimensional complexes.

In a series of studies, Kahle and collaborators~\cite{Kahle2009,Kahle2014,Kahle2014a} investigate another construction of complexes based on the Erd\H{o}s-R\'enyi (ER) model, called \emph{random clique complexes}.
In random clique complexes, a sample from an ER graph is drawn and all complete subgraphs are identified with a face of a simplicial complex.
Kahle characterized the higher-homology groups of random clique complexes~\cite{Kahle2009} (also called flag-complexes~\cite{Kahle2014a}).
Costa and Farber~\cite{Costa2016,Costa2016a,Costa2017,Costa2017a} as well as Fowler~\cite{Fowler2015,Fowler2019} then studied a multiparameter model, first introduced in~\cite{Kahle2014}, that interpolates between the Linial-Meshulam model and the random flag 
complex model.
The model uses an inductive process that starts with an ER graph with~$\maxdim$ nodes and edge-connection (1-simplices) probability~$p_1$. 
Random 2-simplices are then added to the resulting 1-skeleton with probability~$p_2$.
Inductively, random 3-simplices are then added to the resulting 2-skeleton etc.
The above mentioned models can then be recovered for particular settings of~$p_i$~\cite{Costa2016,Costa2016a,Costa2017,Costa2017a,Fowler2015,Fowler2019}, for instance the original Linial-Meshulam model~\cite{Linial2006} can be recovered for $p_1=1$ (all edges are present) and $p_2 = p$ (2-simplices are added to the full graph with probability~$p$).
It can be shown that the multiparameter model has a dual model~\cite{Farber2019,Farber2020} in which one starts with~$\maxdim$ vertices, selects subsets of size~$k$ with probability~$p_k$ (i.e., one constructs a general hypergraph) and then adds all their faces to obtain a random complex.
A model based on such a downward closure of a hypergraph is also analyzed in~\cite{Cooley2020}, where the vanishing of the cohomology groups are characterized in more detail.

\paragraph{Random geometric simplicial complexes}
Similar to how the random abstract simplicial complex models can be seen as extensions to well-known random graph models, there exist extensions of random \emph{geometric} random graph models to simplicial complexes.
Rather than providing an exhaustive discussion our goal here is again to provide pointers to a few key papers and ideas.
For a more comprehensive recent review the interested reader may consult the surveys by Kahle, and Bobrowski and Kahle~\cite{Kahle2011,Bobrowski2018} (see also~\cite{Battiston2020}).

Random geometric models for complexes are in particular relevant as null models for topological data analysis (see~\Cref{sec:Structure}).
In particular, such models enable us to understand various statistics arising in the construction of simplicial complexes from random geometric data such as Euler characteristics~\cite{Ferraz2011}, homology~\cite{Bobrowski2017}, or other statistics of persistence diagrams that arise in applying persistent homology to random geometric data~\cite{Fasy2014}.

\paragraph{Maximum entropy and latent variable models for simplicial complexes}
The above described models are highly stylized models that allow for theoretical explorations.
For null-model hypothesis testing of simplicial data, however, we may one find a model that fixes certain observed statistics, while randomizing all other features of the data.
Formally, this idea can be implemented via maximum entropy models for simplicial complexes, such as the configuration model~\cite{Courtney2016,Young2017} (that keeps the observed (simplicial) degree sequence of the observed data) or exponential random simplex models~\cite{Zuev2015} (see the following section for a definition of exponential random graph and hypergraph models).
So far these models have been explored mostly computationally and far less is known about their theoretical properties.
Apart from maximum entropy models, another important class of random simplicial complex models are latent variable models.
For instance, in the case of random geometric complexes, given some observed simplicial data, we may want to infer a compatible latent geometry for this data.
The resulting geometry may then be interpreted a form of hidden cause for the observed simplicial data.
Outside the realm of geometric models, the study of latent variable models has received little attention for modeling simplicial complexes (see also~\cite{Battiston2020}), but has been mostly restricted to general hypergraphs.

\subsubsection{Probabilistic models for hypergraphs} \label{subsec:prob-models-hypergraphs}

In line with the study of random simplicial and cell-complexes, the study of probabilistic models for general random hypergraphs is still quite young.  
While, in theory, several classes of models are flexible enough to model a wide variety of effects, in practice, only a few specific probabilistic models for hypergraphs have been studied in detail. 
Most developed statistical models for hypergraphs are simple in some way, and thus, tend to be used as null models, as with the random simplicial complex models described in the previous section. 
It should be noted, however, that several models for hypergraphs, especially those that are extensions of exponential random graph models (ERGMs), have their roots in categorical data analysis, an area of statistics dealing with discrete data sets, and thus benefit from some existing methods for parameter estimation and goodness-of-fit testing.  
This gives us some tools for handling higher-order relational data.  
Here we describe several probabilistic models for hypergraphs, starting with generalizations of Erd\H{o}s-R\'enyi random graphs, moving to configuration models that were alluded to in the previous section.
{Finally we discuss exponential random graph models, a family that includes the~$\beta$ model for hypergraphs, and latent variable models such as hypergraph stochastic block models.}

\paragraph{Erd\H{o}s-R\'enyi hypergraphs}  Similar to simplicial complexes, it is natural to generalize Erd\H{o}s-R\'enyi (ER) random graphs to uniform hypergraphs.
Erd\H{o}s-R\'enyi random graphs come in two main forms, the~$G(\maxdim,\maxedg)$ model, where~$\maxedg$ edges are chosen uniformly from the set of all possible $\maxdim \choose 2$ edges on~$\maxdim$ vertices, and the $G(\maxdim,p)$ model, where each edge on~$\maxdim$ vertices is chosen with probability~$p$, which is also described in~\Cref{subsec:randomsimplicialcomplexes}.  Both generalizations of the~$G(\maxdim,\maxedg)$ and~$G(\maxdim,p)$ models, referred to as~$H^d(\maxdim,\maxedg)$ and~$H^d(\maxdim,p)$ models, respectively, have been suggested and explored.  
For the model $H^d(\maxdim,\maxedg)$, the hypergraph generalization of~$G(\maxdim,\maxedg)$, a hypergraph is constructed by choosing $m$~hyperedges from the set of all $\maxdim \choose d$ $d$-uniform hyperedges on~$\maxdim$ vertices. 
Just as asymptotics for ER graphs as $\maxdim \to \infty$ have been studied, similar studies have appeared for the $H^d(\maxdim,\maxedg)$ model. 
For example, in the mid-80s, Schmidt-Pruzan and Shamir showed that if $b \geq 2$ and $\maxedg = c\maxdim$ with $c < 1/b(b-1)$ then asymptotically almost surely the largest component is of order $\log \maxdim$ ~\cite{schmidt1985component}. 
Following-up on~\cite{schmidt1985component}, Karo{\'n}ski and {\L}uczakin showed that phase transition from many small components to single large component occurs when $\maxedg = \maxdim/d(d-1) + O(\maxdim^{2/3})$~\cite{karonski2002phase}. Meanwhile, for the $H^d(\maxdim,p)$ model, the hypergraph generalization of $G(\maxdim,p)$ where each $d$-uniform hyperedge on~$\maxdim$ vertices is selected with probability $p$, thresholds for different Hamiltonian properties have been investigated~\cite{balhamiltonian, clemens2020dirac, dudek2013tight, nenadov2019powers}.  The ER hypergraph model~$H^d(\maxdim,p)$ is an ERGM model, and we will say more about tools for goodness-of-fit and hypothesis testing below.

\paragraph{Configuration models}
Erd\H{o}s-R\'enyi hypergraph models give rise to distributions over the space of all $d$ uniform hypergraphs with~$\maxdim$ vertices.
Configuration models give rise to distributions over the set of hypergraphs with a fixed edge size sequence $\mathbf{k} = ( k_e \ | \ e \in \Ed)$ and degree sequence $\mathbf{d} = (d_v \ | \ v \in \Ve)$. 
{As a concrete example,} in a stub-labeled configuration model, as described in~\cite{ghoshal2009random} and~\cite{chodrow2019}, a graph or hypergraph is constructed by associating~$d_v$ ``stubs'' to the vertex $v$ and then connecting $k_e$ stubs to form each edge in~$\Ed$.  
{We remark that this process may result in hypergraphs with \emph{degenerate} edges, i.e., edges that contain multiple copies of a single vertex, as well as multi-edges, i.e., multiple edges connecting the same nodes.
However, conditioning on non-degeneracy and simplicity (no multi-edges), every non-degenerate stub-labeled hypergraph with edge sizes $\mathbf{k} = ( k_e \ | \ e \in \Ed)$ and degree sequence $\mathbf{d} = (d_v \ | \ v \in \Ve)$  is equally likely to be produced under the stub-matching algorithm. 
Similar considerations can be applied to configuration models which allow for multi-edges and/or self-loops~\cite{chodrow2019}.
Rather than applying the above mentioned stub-matching process to draw samples from a configuration model, in practice, Markov chain Monte Carlo (MCMC) techniques are typically used to create samples.
These MCMC techniques can be easily adjusted to work for both stub-labeled configuration models as exemplified above, and vertex labeled configuration models (see~\cite{chodrow2019,fosdick2018configuring} for more detailed discussions).
}

Both~\cite{ghoshal2009random} and~\cite{chodrow2019} illustrate the usefulness of {configuration models} as null models for hypergraphs, the former focusing on tripartite hypergraphs describing data from an online photo-sharing website and the latter focusing on a variety of collaboration and communication networks.  
Recently, Chodrow and Mellor~\cite{chodrow2020annotated} extended configuration models to \emph{annotated hypergraphs}, generalizations of directed hypergraphs where each vertex has a ``role'' in each edge.

\paragraph{Exponential random graph models} Perhaps the most flexible family of probabilistic models for hypergraphs are exponential families and can be described as extensions of exponential random graph models.  This framework is especially helpful for analysis, since if the model is log-linear, then tools from categorical data analysis can be used.

An exponential family random graph model (ERGM) is a collection of probability distributions on the space of all graphs~$\mathfrak G_\maxdim$ with~$\maxdim$ vertices such that the probability for $\mathcal G\in\mathfrak G_\maxdim$ is
\begin{equation}\label{eq:ERGM}
P_{\theta}(\mathcal G) = Z(\theta) e^{{\theta} \cdot t(\mathcal G)}, 
\end{equation}
where~$G$ is represented as a vector in~$\R^{\maxdim^2}$ obtained by flattening the adjacency matrix of~$G$, 
$\theta$~is a row vector of parameters of length~$q$, the map $t: \R^{\maxdim^2} \to \R^q$ 
computes the sufficient statistic, and~$Z(\theta)$ is a normalizing constant.  
The image of the sufficient statistic map~$t$ is a vector in which each entry is a network statistic used to specify the model, such as edge count, degree of a given vertex, triangle count, etc.

Exponential random graph models have been extended to hypergraphs in two ways.  
The first method considers a 
{hypergraph as a \emph{two-mode network} or \emph{incidence graph}}\footnote{See for example~\cite{Walsh1975}. The incidence digraph is also called the Levi digraph or K\"onig digraph.}.
More specifically, any hypergraph~$\mathcal H$ with $\maxdim$~vertices and $\maxedg$~hyperedges
can be identified with a bipartite graph~$\mathcal{B}_\Hg$ with $\maxdim+\maxedg$ vertices of two types: The vertices $v_1, \dotsc, v_\maxdim$ correspond to the vertices of~$\Hg$, the vertices $e_1, \dotsc, e_\maxedg$ correspond to the hyperedges of~$\Hg$ and there is an edge in $\mathcal{D}_\Hg$ between vertices~$v_i$ and~$e_j$ of~$\mathcal{B}_\Hg$ if and only if vertex~$v_i$ of~$\Hg$ is contained in hyperedge~$e_j$ of~$\Hg$.
Several ERGMs for two-mode networks have been suggested and explored in~\cite{skvoretz1999logit,pattison2004building,wang2009exponential} and the {\tt ergm} package for {\tt R}~\cite{hunter2008ergm} has some functionalities for handling two-mode networks.

The second type of extension builds on the observation that an ERGM can easily be enlarged to collections of probability distributions on the space of all hypergraphs with~$\maxdim$ vertices.  This can be done by using an appropriate vector representation of a hypergraph and a meaningful sufficient statistic~$t$, such as the degree sequence. The ER hypergraph model~$H^d(\maxdim,p)$ is an ERGM model when~$t$ is the edge count.

Although we can describe an ERGM for any set of sufficient statistics of a hypergraph data set, only a few specific ERGMs have been studied in detail and thus suitable for practical use. One example is the $\beta$~model for hypergraphs, an ERGM for non-uniform hypergraphs where the sufficient statistic is the degree sequence. Algorithms for fitting parameters and problems related to the existence of MLE discussed are described in~\cite{betaHypergraphs}.
In~\cite{betaHypergraphs}, the authors give three hypergraph variations for the $\beta$~model: one for uniform hypergraphs, one for layered hypergraphs, and one for general hypergraphs.  
We would call each of these log-linear along the lines of~\cite[Definition 3.1]{gross2019algebraic}.  
The log-linear designation signifies that the sufficient statistic is a linear function on a reasonable contingency table representation of the hypergraph with possible structural zeros.  
Indeed, we can always represent a $k$-uniform hypergraph on~$\maxdim$ vertices as a $\maxdim \times \dotsb \times \maxdim$ dimensional $k$-way table with structural zeros on the $(i_1, \ldots, i_k)$ entry whenever $i_j = i_{\ell}$ for one pair $j \neq \ell$. 
Thus, we can view a hypergraph as a $0$-$1$ contingency table with structural zeros.
This viewpoint opens up the field of categorical data analysis and all log-linear models to hypergraph data and allows the use of algebraic statistics for goodness-of-fit testing as in~\cite{GPS16, karwa2017statistical, gross2021random}.

\paragraph{{Latent variable models and extensions of stochastic blockmodels}}
Finally, another log-linear ERGM extension for hypergraphs are \emph{hypergraph stochastic block models}~\cite{Ghoshdastidar2014, lesieur2017statistical, Kim2017}, which are generalizations of stochastic block models for graphs. 
In the simplest hypergraph stochastic block model variant, vertices are partitioned into $k$~clusters and $d$-uniform hypergraphs are constructed by choosing each size~$d$ edge with vertices all contained in a single cluster with some fixed probability~$p$ and choosing every other size~$d$ hyperedge with fixed probability~$q$, generally with $p > q$.  
The sufficient statistic for this model is a pair of two numbers: the number of within-cluster hyperedges and the number of in-between-cluster hyperedges.  

While hypergraph stochastic block models can be used to test for homophily effects in a hypergraph dataset when the cluster membership of each vertex is known, one of their main applications, as described in the next section, is for community detection and clustering when the cluster membership of each vertex is unknown. 
{In this case, the models become \emph{latent variable models} and the goal becomes trying to determine the most likely membership assignment of each node.  For graphs, inference is usually done using coordinate ascent methods, such as variants of the expectation-maximization  algorithm,  Monte Carlo Markov chain methods, and variational methods (and overview of such methods can be found in \cite{lee2019review}); these methods are just now starting to be developed for hypergraphs \cite{ke2019community}.

The simplest forms of hypergraph stochastic block models tend to generate graphs where all the nodes in the same community have similar degrees, however, for applications, it is desirable to have a model where the node degrees can display some heterogeneity. In the graph setting, stochastic block models that also allow for degree heterogenity are referred to as \emph{degree-corrected stochastic block models}.  In the exponential family setting, the sufficient statistics of these models are obtained by taking the sufficient statistic of the corresponding stochastic block model and appending the degree sequence of the vertices~\cite{yang2015social}. Moving from graphs to hypergraphs, similar variants of hypergraph stochastic block models that allow for degree heterogenity have been proposed. For example, in~\cite{ke2019community}, Ke, Shi, and Xia define \emph{degree-corrected hypergraph stochastic block models}, and develop a method for community detection using tensor decomposition.
Furthermore, degree-corrected hypergraph stochastic block models form the basis of the algorithms for clustering described by Chodrow, Veldt, and Benson in~\cite{chodrow2021generative} where they illustrate their usefulness for detecting ground truth communities through several datasets. It should be noted, that the models described in~\cite{chodrow2021generative} allow for heterogenity in the edge size as well, meaning, that in this case, the models are collections of probability distributions over hypergraphs with varying edge size.}

\subsection{Analyzing higher-order relational data: application examples} \label{ssec:stat_examples}
Let us now consider some example applications, which are based on the analysis of relational data (from a statistical point of view).

Applications in which hypergraph-based techniques have been considered to analyze relational data include anomaly detection~\cite{Silva2008}, population stratification~\cite{Vazquez2008}, as well as the analysis of folksonomies~\cite{Zhang2010,ghoshal2009random}.
We mention again the work of Chodrow et al.~\cite{chodrow2019,chodrow2020annotated} on hypergraph configuration models, which can serve as null models for various statistical applications.

One of the most popular analyses of relational data in the network setting is community detection (or network clustering), in which one aims to partition the set of vertices in the system into groups such that these groups are more similar to each other than to rest of the network.
A large number of methods for community detection exist for networks~\cite{fortunato2016community} based on various notions of what constitutes a good community~\cite{schaub2017many}.
Unsurprisingly, the problem of community detection has been considered as well for higher-order relational data.
For instance, Vazquez~\cite{Vazquez2009} presented some early work on detecting hypergraph communities using a Bayesian framework. 
Some more theoretical work includes Ghoshdastidar et al.~\cite{Ghoshdastidar2017,Ghoshdastidar2014}, who analysed spectral clustering and its consistency for hypergraphs under {a planted partition model, a specific type of stochastic block model}.
Kim et al.~\cite{Kim2017} considered sum-of-squares approaches for similar hypergraph stochastic block models, and Chien et al.~\cite{Chien2018} have analyzed the optimal rates of community detection for $d$-wise hypergraph stochastic block models, where the probability of a hyperedge is dependent on the distribution of the vertex block assignments of the vertices contained in the hyperedge. {Moreover, community detection for higher-order relational data has motivated much work in spectral hypergraph theory~\cite{angelini2015spectral, chan2018spectral, chan2020generalizing, li2018submodular, tudisco2021node,zhou2006learning} and hypergraph cuts and splitting functions~\cite{hein2013total, veldt2020hypergraph, yaros2013imbalanced}.}

As higher-order relational data may also be abstracted via simplicial complexes, there are also a number of works that use such a modeling perspective, often blending tools from topological data analysis (\Cref{sec:Structure}) with statistical tools.
In particular, there has been work on generalizing the idea of triadic closure in networks---the tendency of two nodes to connect, if they have a mutual neighbor---to higher-order networks~\cite{Benson2018,Patania2017}.
While these works have been concerned with how the topology of (higher)-order networks evolve, there are also higher-order models for analyzing edge-data on networks that utilize the Hodge decomposition for the purpose of extracting aggregating rankings in a consistent way~\cite{Jiang2011} or for smoothing and filtering flows and trajectories on networks and simplicial complexes~\cite{barbarossa2016introduction,Schaub2018,Barbarossa2018,Jia2019,Schaub2020}.

\section{Network Dynamical Systems with Higher-Order Interactions}\label{sec:Dynamics}

{Network dynamical systems describe the joint evolution of interacting dynamical nodes. The coupling between the nodes can lead to intriguing collective network dynamics such as synchronization, where nodes evolve in unison. 
A key question is how the network interactions---both the network structure as well as the functional form of the coupling---shape such collective dynamics.
While graphs have traditionally been used to encode network structure, we focus on more general, ``higher-order'' approaches to capture the coupling structure.
Naturally, this yields questions such as: What higher-order structures are appropriate for network dynamical systems? And how do higher-order interactions shape the dynamics? Here we focus on network dynamical systems where the nodes/vertices are dynamical units, even though (higher-order) networks with dynamical (hyper)edges have also recently been discussed~\cite{Millan2019a,Nijholt2022}.

Of course, network dynamical systems can yield data (e.g., by sampling from the trajectory) to which methods outlined in the previous sections can be applied to. However, here we will take a more general dynamical systems approach to elucidate general properties of interacting dynamical systems.
}

\subsection{Network dynamical systems with pairwise interactions}
Traditionally network interactions between nodes are encoded in a directed, weighted graph~$\G=(\Ve,\Ed)$ on $\maxdim$~vertices $\Ve=\sset{1, \dotsc, \maxdim}$ with weighted adjacency matrix~$\mathbf{A}$.
For simplicity, let us suppose that we have~$\maxdim$ identical nodes with dynamics of the form $\dot x_k = F(x_k)$, where $x_k\in\Rd$ is the \emph{state vector} of node~$k\in\Ve$ and $F:\Rd\to\Rd$ is a function that describes the intrinsic coupling of the state vector of the node.
Suppose further that any interaction between nodes can be described by a pairwise coupling function $G:\Rd\times\Rd\to \Rd$, which describes how the states of two connected nodes in~$\G$ interact.
The tuple $(\G,F,G)$ defines a {network dynamical system} through the set of differential equations
\begin{equation}\label{eq:NetDynPairs}
    \dot x_k = F(x_k) + \sum_{j=1}^\maxdim A_{jk}G(x_k,x_j)
\end{equation}
for $k = 1,\dotsc,\maxdim$.
While discrete time network dynamics of a similar form are {certainly also of interest~\cite{Chazottes2005}}, for simplicity, we will consider primarily continuous time dynamics in the following exposition.

{The \emph{collective network dynamics} are now determined by the evolution of the joint state of all nodes ${x=(x_1, \dotsc, x_\maxdim)}$ through~\eqref{eq:NetDynPairs}.}
Specifically, the {collective network dynamics} are determined by (i)~the structure of the graph $\G$, as encoded in the adjacency matrix~$\bm A$, (ii)~the intrinsic dynamics~$F$ of each node, and (iii)~the pairwise coupling function~$G$~\cite{Stankovski2017}.
This setup gives rise to many classical problems in network dynamical systems.
On the one hand, how do the collective network dynamics~$x(t)$ depend {on~$(\G, F, G)$ that determine the network dynamical system}?
For example, how are these dynamics perturbed as edges are added or removed?
{How do properties of the network structure (e.g., modularity~\cite{Bick2018c,lambiotte2021modularity}) influence a dynamical process on a network?}
On the other hand, how can the network structure and interactions be inferred from measurements of the dynamics~$x(t)$? This last problem is commonly known {as inference of network dynamical systems}~\cite{Timme2014}.

It is important to note that network dynamical systems described via~\cref{eq:NetDynPairs} have \emph{additive interactions}~\cite{Bick2015}.
Specifically the interactions are in general nonlinear in the state variables~$x_k$, but \emph{linear} in the coupling weights~$A_{kj}$.
Hence, changes in the interaction graph~$\G$ can be naturally incorporated by adjusting the value of~$A_{kj}$.
While this setup~\cref{eq:NetDynPairs} is arguably one of the most commonly considered models for network dynamical systems, in general this linearity in the coupling weights might not hold. 
Accordingly, we may need to go beyond pairwise couplings.

\subsection{Network dynamical systems with higher-order interactions}\label{sec:Nonadditive}
For general network dynamical systems, the state evolution of each node may not be expressible as a superposition of pairwise interactions as in~\cref{eq:NetDynPairs}. 
For instance, suppose that the dynamics evolve according to
\begin{equation}\label{eq:NetDynNonlin}
    \dot x_k = F(x_k)+H_k(x)
\end{equation}
for $k=1, \dotsc,\maxdim$ where each $H_k:\R^{d\maxdim}\to\Rd$ determines the influence of the joint state of the network~$x=(x_1, \dotsc, x_\maxdim)$ on node~$k$.
Clearly, the function~$H_k$ may depend not only on two node states, but may involve multiple nodes concurrently.

Network dynamical systems of the form~\cref{eq:NetDynNonlin} have been considered as \emph{coupled cell systems}~\cite{Stewart2003}.
{Fix a directed graph~$\G_\mathrm{CSS}$ with~$\maxdim$ vertices that encodes the node dependencies. Consider now all functions~$H = (H_1, \dotsc, H_\maxdim)$ that are compatible with~$\G_\mathrm{CSS}$ in the sense that~$H_k$ depends nontrivially on~$x_j$ if there is an edge from node~$j$ to node~$k$ in the edge set~$\Ed(\G_\mathrm{CSS})$. Note that despite node dependencies being captured by a graph, this does not exclude the possibility of nonlinear interactions involving three or more nodes as in network dynamical systems of the form~\cref{eq:NetDynPairs}.
However, the main goal of the coupled cell system formalism is to elucidate the properties of all dynamical systems~\cref{eq:NetDynNonlin} that are compatible with the network structure~$\G_\mathrm{CSS}$ simultaneously (rather than considering~\eqref{eq:NetDynNonlin} for a specific~$H$):
This yields insights how the generic dynamical behavior, such as bifurcation and synchrony, depend on the imposed network structure (encoded by~$\G_\mathrm{CSS}$); see, for example,~\cite{Aguiar2018,Nijholt2019}.
}

{However, rather than looking at generic properties, one is often interested in the dynamics for specific (classes of) coupling functions~$H$. To this end, }we may expand the general dynamics~\eqref{eq:NetDynNonlin} formally as
\begin{equation}\label{eq:NetDynNonlinExp}
    \dot x_k = F(x_k) + \sum_{j=1}^\maxdim A_{jk}G_k(x_k, x_j) + \sum_{j,l=1}^\maxdim A^{(3)}_{jlk}G_k^{(3)}(x_k, x_j, x_l) + \dotsb,
\end{equation}
where the (formal) adjacency matrix~$\mathbf{A} = (A_{jk})$ together with the coupling functions~$G_k$ characterize the \emph{pairwise} network interactions (cf.~\Cref{eq:NetDynPairs}), and the coefficients of~$A^{(s)}$  and coupling functions~$G_k^{(s)}$ (with $s\ge3$) the \emph{nonpairwise} interactions\footnote{We assume that the~$G^{(s)}_k$, $s>2$, depend nontrivially on all of its arguments.}.
With respect to the formal expansion~\cref{eq:NetDynNonlinExp} the terms~$G^{(s)}_k$ with coefficients~$A^{(s)}_{j_1\dotsb j_s}$ may thus be called \emph{higher-order network interactions}. 
For example, $A^{(3)}_{jlk}$ and~$G_k^{(3)}(x_k, x_j, x_l)$ describe the joint influence of nodes~$l,j$ on node~$k$ in the expansion~\cref{eq:NetDynNonlinExp}.

Analogously to how we associate a graph~$\G$ to the pairwise dynamics~\cref{eq:NetDynPairs} by interpreting~$\mathbf{A}$ as an adjacency matrix, we may want to associate an appropriate (combinatorial) mathematical structure to the higher-order interaction terms.
Similar to the graph case, we could then analyze this mathematical object and hopefully elucidate some interesting {dynamical} properties of higher-order system~\cref{eq:NetDynNonlinExp}.
This reasoning motivates a number of questions: 
\begin{enumerate}[label=\textnormal{(Q\arabic*)}]
    \item\label{Q:Dyn} What are dynamical consequences of higher-order network interactions?
    \item\label{Q:Model} Are higher-order interactions crucial to understand dynamics of real world systems?
    \item\label{Q:Struct} Is there an appropriate structure (hypergraph, simplicial complex, etc.) to represent a generically coupled network dynamical system such as~\eqref{eq:NetDynNonlin}?        
\end{enumerate}

    Intuitively, as~\Cref{eq:NetDynNonlinExp} exhibits a more general class of vector fields on the right hand side compared to~\Cref{eq:NetDynPairs}, we may expect that there will be a number of new phenomena higher-order networks can exhibit.
    Accordingly, we may argue that these higher-order interactions can indeed be crucial, if a real-world system can be represented only in terms of higher-order interactions~\cref{eq:NetDynNonlinExp}.
    A somewhat more careful inspection of our above example shows  however that the issue is somewhat more subtle.
    For instance, for the coupled cell systems~\cref{eq:NetDynPairs}, there is a meaningful \emph{graph-based} description in terms of~$\G_\mathrm{CSS}$ associated to the system, despite the fact that generally the system will be a network dynamical system with higher-order interactions. 
    Furthermore, the expansion~\cref{eq:NetDynNonlinExp} will in general not be unique, and we can associate many possible higher-order networks to a system of the form~\cref{eq:NetDynNonlin}.
    Indeed, we believe that questions~\ref{Q:Model} and~\ref{Q:Struct} do not have well-defined answers, and we will discuss in \Cref{ssec:dynamics_hypergraphs_and_scs} how choosing an appropriate structure can be a matter of perspective.
    Before returning to these conceptual matters, {we will first} explore question~\ref{Q:Dyn} and some aspects of~\ref{Q:Model} more concretely and outline some partial answers that have been given so far in the literature.

\subsection{Effects of higher-order interactions in network dynamical systems}

In the following, we will discuss some consequences of higher-order network interactions on the dynamics and why it may be crucial to account for these (cf.~\ref{Q:Dyn} and~\ref{Q:Model} in the previous section).
We focus on specific examples relevant in a variety of real-world dynamical systems rather than attempting to give a comprehensive overview; see~\cite{Battiston2020} for a far more extensive list of references.

\subsubsection{Phase oscillator networks}\label{sec:PhaseOscHigherOrder}
Networks of simple phase oscillators have received tremendous attention in recent years.
These networks yield insights into a range of synchronization phenomena, a prominent form of collective dynamics in which distinct nodes in a network evolve in unison.
Reviews on synchronization phenomena include~\cite{Acebron2005,Rodrigues2016,Bick2018c,Omelchenko2018}. {Probably the most famous example of a phase oscillator network is the Kuramoto model~\cite{Kuramoto,Strogatz2000} and its variations, where the state of the oscillator at node~$k$ is given by a phase $\theta_k\in \Tor:=\mathbb{R} / 2\pi\mathbb{Z}$.
For the Kuramoto model on an arbitrary graph~$\G$ with weighted adjacency matrix~$\bm A = (A_{jk})$, the phase variable of node~$k$ evolves according to
\begin{align}\label{eq:KuramotoSakaguchiNet}
    \dot\theta_k &= \omega_k + \frac{K}{\maxdim}\sum_{j=1}^\maxdim A_{jk}\sin(\theta_j-\theta_k),
\end{align}
where $\omega_k\in\R$ are the intrinsic frequencies of the oscillators and interactions are pairwise along edges (the equations are of the form~\eqref{eq:NetDynPairs}).
If~$\G$ is the complete graph and the~$\omega_k$ are independently sampled from a unimodal probability distribution, we recover Kuramoto's original equations that can be analyzed explicitly~\cite{Bick2018c}.}

This setup has been recently generalized by adding higher-order interaction terms to the standard Kuramoto model. 
For instance, Skardal and Arenas~\cite{Skardal2019a} consider networks of~$\maxdim$ phase oscillators with ``simplicial'' coupling, where oscillator~$k$ evolves according to
\begin{equation}\label{eq:KuramotoSakaguchiHO}
\begin{split}
    \dot\theta_k &= \omega_k + \frac{K}{\maxdim^2}\sum_{j\neq k} \left(2\sin(\theta_j-\theta_k)+\sin(2\theta_j-2\theta_k)\right)\\
                 &\qquad\qquad+ \frac{K}{\maxdim^2}\sum_{j\neq k \neq l} \sin(\theta_j+\theta_l-2\theta_k)
                 \end{split}
\end{equation}
where the last sum is over pairwise distinct indices~$j,l,k$. 
Note that already the pairwise interactions in this model are all-to-all but distinct from the Kuramoto model.
Specifically, the first sum in~\cref{eq:KuramotoSakaguchiHO} (over the pairwise couplings) includes first and second harmonics and, moreover, the second sum incorporates nonlinear higher-order terms (i.e., $A_{jlk}^{(3)}\neq0$ for some~$j,l,k$ in~\eqref{eq:NetDynNonlinExp}). 
We remark that such interaction terms had been previously considered in the context of nonlinear interactions~\cite{Komarov2015}.

These and other higher-order terms in the Kuramoto model can have a range of dynamical effects and are directly relevant to understand the dynamics of oscillations since they arise in a number of physical systems. Examples include oscillatory dynamics in electronic circuits~\cite{Heger2016}, the dynamics of coupled nanomechanical oscillators~\cite{Matheny2019a}, and optical devices~\cite{Pelka2019}.
Apart from an impact on synchronization~\cite{Komarov2015,Skardal2019a,lucas2020multiorder}, higher-order interactions can stabilize splay/twisted solutions~\cite{Bick2022},
lead to multistability~\cite{Tanaka2011a}, 
heteroclinic cycles~\cite{Bick2018,Bick2018a,Bick2017c}, 
and chaotic dynamics~\cite{Bick2016b,Leon2022}. 
Other generalizations of the Kuramoto model to simplicial complexes~\cite{Millan2019a} are possible that lead to ``explosive'' synchronization phenomena---an effect that one may expect generically when generalizing a classical model~\cite{Kuehn2020}.

\subsubsection{Ecological networks}

Lotka--Volterra-type equations provide a classical model for the dynamics of species populations. Interactions between species in these equations are pairwise. It has been argued that incorporating higher-order interactions are essential for the mechanisms that lead to coexistence of species that emerge in diverse competitive networks~\cite{Abrams1983, Levine2017}.
For example, Allesina and Levine~\cite{Allesina2011} consider networks of multiple competitively interacting species, where the relative abundance~$p_k$ of species~$k$ evolves according to 
\begin{align}
{\dot p_k} &= \sum_{j=1}^{\maxdim}K_{jk}p_jp_k - \sum_{j,l=1}^\maxdim K_{lj}p_lp_jp_k
\end{align}
subject to $\sum_{j=1}^\maxdim p_j = 1$. This is a network dynamical system with {interactions of order two and three} in terms of the expansion~\eqref{eq:NetDynNonlinExp}.
Including higher-order interactions can make {species coexistence robust to the perturbation of both population abundance as well as parameter values~\cite{grilli2017higher}}.

\subsubsection{Neuroscience}
\label{sec:Neurononadd}
Nonadditive higher-order interactions also play a functional role in the dynamics of neural networks (see, e.g., Refs.~\cite{Ariav2003,Jahnke2015} and references therein).
To understand the implications of higher-order interactions for the collective network dynamics, Memmesheimer and Timme~\cite{Memmesheimer2012} consider neural networks with nonlinear amplification and saturation in the input (the dendrites) of each neuron. 
Specifically, they consider leaky integrate-and-fire neurons with a non-additive input function~$\sigma$.
If the input for a given neuron is low then the inputs add linearly.
Somewhat larger inputs are amplified superlinearly, before saturating to an overall maximal excitation to each neuron.
While the coupling considered in~\cite{Memmesheimer2012} describes noncontinuous dynamics, the coupling is nonadditive as in~\eqref{eq:NetDynNonlinExp} and the input to each neuron depends nonlinearly on the joint state of all its input neurons that emit a spike. 
In addition to propagation of synchronous activity~\cite{Memmesheimer2012}, these nonadditive interactions allow for the emergence of high-frequency oscillations~\cite{Memmesheimer2010} and memory processes~\cite{Jahnke2015}.

\subsubsection{Contagion, diffusion, and other network dynamics with higher-order interactions}\label{ssec:discrete_time_dynamics}
While we only consider network dynamical systems determined by ordinary differential equations, higher-order interactions also arise in discrete time dynamical system or systems with a discrete state space.
This includes for example contagion dynamics on a simplex~\cite{Iacopini2018,de2020social,bodo2016sis,Higham2021}, where the contagion of one node depends on the joint state of two or more nodes, or adaptive voter models with non pairwise interactions and rewiring~\cite{Horstmeyer2019}. 
One possible approach to gain analytical insights into these dynamics are mean-field approaches that capture the higher-order interactions: For example, these can uncover how critical transitions in such systems change from subcritical to supercritical and vice versa~\cite{Kuehn2020}.
Other examples of higher-order dynamical processes include opinion formation and consensus processes on hypergraphs~\cite{DeVille2020,neuhauser2020multibody,neuhauser2020opinion,Hickok2021}, diffusion processes~\cite{carletti2020dynamical,carletti2020random,Schaub2020,Reitz2020,Millan2021} (see also~\Cref{sec:Hodge}), and replicator dynamics~\cite{Alvarez-Rodriguez2021}.

\subsection{Algebraic structures for network dynamics with higher-order interactions}
Within this section we have thus far considered higher-order interactions from the perspective of a (formal) expansion of the vector field of a network dynamical system.
As we discussed in regards to standard network dynamical systems with pairwise interactions, the network structure can be seen as a graph whose vertices correspond to the individual dynamical units and the edges encode interactions. 
By extension, this suggests that network dynamics with higher-order interactions {should be interpreted as} an appropriate extension of a graph such as a simplicial complex or a hypergraph:
Specifically, whereas the interactions in~\eqref{eq:NetDynNonlinExp} that are mediated by the adjacency matrix~${\bf A}$ are typically interpreted as edges ($1$-simplices) of a graph, the coupling terms $A^{(s)}$ in~\Cref{eq:NetDynNonlinExp} would then correspond to higher-dimensional simplices and/or hyperedges.
In the following we will discuss the relationship between network dynamical systems with higher-order interactions~\eqref{eq:NetDynNonlinExp} and their mathematical representations discussed in \Cref{sec:Prelims}.

\newcommand{\abs}[1]{\left|#1\right|}

\subsubsection{Network dynamics on hypergraphs and simplicial complexes}\label{ssec:dynamics_hypergraphs_and_scs}
Given a simplicial complex or hypergraph, there are many ways to define a dynamical system on it. As mentioned above, Skardal and Arenas consider a generalization of the Kuramoto model on a simplicial complex (containing all possible 2-simplices) by~\eqref{eq:KuramotoSakaguchiHO}. Other generalizations of the Kuramoto model are possible~\cite{Millan2019a}.
For a hypergraph~$\Hg$ write $\Ed_k := \set{e\in \Ed(\Hg)}{k\in e}$ for the set of edges that are incident to node~$k$. 
For a point $x\in\R^{d\maxdim}$ and a hyperedge $e = \sset{j_1< \dotsb < j_Q}$ write $x_e = (x_{j_1}, \dotsc, x_{j_Q})$ for the projection of~$x$ on the coordinates contained in~$e$.
In this more general setting, Mulas and colleagues~\cite{Mulas2020} derive a master stability approach for dynamical systems on a hypergraph~$\Hg$ with identical nodes whose state evolves according to
\begin{align}\label{eq:DynHG}
\dot x_k = F(x_k) + \sum_{e\in \Ed_k(\Hg)} G_{k;e}(x_e). 
\end{align}
In their setup, few assumptions are made on the coupling functions~$G_{k;e}$: While they have to be invariant under permutations of the arguments, they could be linear in the arguments~$x_j$ and explicitly depend on the node~$k$. Related work includes the analysis of more elaborate synchrony patterns~\cite{Salova2021a,Salova2021,Aguiar2020}.

While one can write down a network dynamical system for a given a hypergraph~$\Hg$, the converse---associating an algebraic structure with a given dynamical system---is not as straightforward; cf.~Question~\ref{Q:Struct} in \Cref{sec:Nonadditive}. In particular, this depends not only on the coefficients~$A^{(s)}$ but also on the assumptions one wants to make on the interaction functions~$G$. The answer to this question depends on the perspective one takes: First, consider the hypergraph~$\Hg$ on~$\maxdim$ nodes with a single edge $e=\sset{1, \dotsc, \maxdim}$. The hypergraph is invariant under any permutations of the vertices. By contrast, the dynamical system~\eqref{eq:DynHG} is not necessarily symmetric (equivariant) with respect to this symmetry operation (unless $G_{k;e}=G_{e}$ for all~$k$). Second, if $G_{k;e}(x_e) = \sum_{q=1}^Q x_{j_q}$ then for~\eqref{eq:DynHG} written as a formal expansion~\eqref{eq:NetDynNonlinExp} we have $A^{(s)}=0$, $s\geq 3$, for any~$\Hg$. In other words, the interactions are additive despite the underlying algebraic structure being a hypergraph.

Finally, note that any hypergraph can be identified with its bipartite incidence graph that captures pairwise incidence relations between nodes and hyperedges (see~\Cref{subsec:prob-models-hypergraphs}). In the context of a recent generalization of the coupled cell framework to higher-order network dynamics~\cite{Aguiar2020}, for patterns of cluster synchrony one gives information about the other the perspectives are not equivalent.

\subsubsection{Hypergraphs and simplicial complexes for network dynamics}
{From the perspective of asymptotic expansions, }the name \emph{higher-order interactions} suggests that the vector field determining the network dynamics can be expanded {in some small parameter}. Indeed, we considered~\eqref{eq:NetDynNonlinExp} as a formal expansion of the generic interactions in~\eqref{eq:NetDynNonlin}. Higher-order network interactions can also arise in asymptotic expansions in a small parameter.
We will illustrate this by considering phase oscillator dynamics as (higher-order) expansions of weakly coupled oscillators. {Specifically, consider~\eqref{eq:NetDynNonlin} with~$H = \eps\tilde H$. As in~\Cref{sec:Nonadditive}, the function~$H$ (or its rescaled version~$\tilde H$) captures the network interactions: Node~$j$ influences node~$k$ if~$H_k$ depends nontrivially on~$x_j$.
Now consider nodes with intrinsic oscillatory dynamics, that is, node~$k$ has state~$x_k\in\R^d$ and $\dot x_k = F(x_k)$ gives rise to an asymptotically stable limit cycle. 
This implies that 
\begin{equation}\label{eq:NetDynNonlinOsc}
    \dot x_k = F(x_k)+\eps \tilde H_k(x)
\end{equation}
has a normally hyperbolic attracting $\maxdim$-dimensional torus for $\eps=0$, which persists for sufficiently small~$\eps>0$~\cite{Fenichel1972,Ashwin1992}; write $(\theta_1, \dotsc, \theta_\maxdim)\in\Tor^\maxdim$ for a point on this torus.}
{This means that if the coupling is sufficiently weak,} the network dynamics can be solely described by the evolution of the phases~$\theta_k$ (as the amplitudes are tied to these phase variables).
A \emph{phase reduction} is an approximation of the dynamics of~$\theta_k$ on the invariant torus obtained through an asymptotic expansion in the small parameter~$\eps$; 
we refer to~\cite{Nakao2015,Ashwin2015,Pietras2019} for and introduction to phase reductions, how to compute them, and how they can help understand network dynamics.
For generic interaction function~$\tilde H_k$, one would expect higher-order interactions in the phase reduction: Ashwin and Rodrigues~\cite{Ashwin2015a} calculated these explicitly for a fully symmetric network of units close to a supercritical Hopf bifurcation to find nonpairwise phase coupling terms already to first order in~$\eps$.

{Important is that even if the nonlinear oscillators are additively coupled---a network dynamical system~\eqref{eq:NetDynPairs} \emph{without} higher-order interactions---its phase dynamics can have nonadditive coupling---a network dynamical system~\eqref{eq:NetDynNonlinExp} \emph{with} higher-order interactions. Such phase dynamics} were termed \emph{effective coupling} in~\cite{Kralemann2014}.
Here, we illustrate this using the example by Le\'on and Paz\'o~\cite{Leon2019a}, who calculated this phase reduction explicitly for a network of globally coupled complex Ginzburg--Landau oscillators~\cite{Hakim1992}{; the underlying network is the complete graph on~$\maxdim$ vertices.}
Specifically, the state of oscillator~$k$ is given by $z_k\in\C$ and evolves according to
\begin{equation}\label{eq:CGLEq}
    \dot z_k=z_k-(1+\mathrm{i}c_2)|z_k|^2z_k+\eps(1+\mathrm{i}c_1)\left(\frac{1}{\maxdim}\sum_{j=1}^\maxdim(z_j-z_k)\right),
\end{equation}
where the coupling is diffusive and pairwise and~$c_1, c_2$ are real parameters and $\mathrm{i}:=\sqrt{-1}${---this is a network dynamical system~\eqref{eq:NetDynPairs} {without} higher-order interactions}. 
The phase dynamics of~\eqref{eq:CGLEq} up to second order in~$\eps$ are given by
\begin{equation}\label{eq:PhaseCGLEq}
\begin{split}
    \dot\theta_k &= \omega + \eps\frac{\eta}{\maxdim}\sum_{j=1}^\maxdim \sin(\theta_j-\theta_k+\alpha)
    +\eps^2\frac{\eta^2}{4}\bigg(
    \frac{1}{\maxdim}\sum_{j=1}^\maxdim \sin(\theta_j-\theta_k+\beta)\\
    &\qquad\quad+\frac{1}{\maxdim^2}\sum_{j,l=1}^\maxdim\sin(2\theta_j-\theta_l-\theta_k)
    -\frac{1}{\maxdim^2}\sum_{j,l=1}^\maxdim\sin(\theta_j-\theta_l-2\theta_k+\beta)
    \bigg),
\end{split}
\end{equation}
where $\omega = -c_2+\eps(c_2-c_1)$, $\beta=\arg(1-c_1^2+2c_1i)$, $\eta = \sqrt{(1+c_1^2)(1+c_2^2)}$ yield the relationship between the parameters of the phase equations and the parameters in~\eqref{eq:CGLEq}.
Note that the first-order phase dynamics are described by the Kuramoto equations~\eqref{eq:KuramotoSakaguchiNet} for identical oscillators on a complete graph with an additional phase-shift parameter{~$\alpha$}.
The second-order terms now yield higher harmonics in the pairwise interactions as well as nonadditive triplet interactions. 
{That means that the phase-reduced system~\eqref{eq:PhaseCGLEq} has higher-order interactions in terms of the formal expansion~\eqref{eq:NetDynNonlinExp}, while the original system~\eqref{eq:CGLEq} did not.}
In this example, the oscillator coupling was all-to-all but the same observation is expected if interaction was along a noncomplete graph: While the phase evolution of a given node is directly influenced by its neighbors, it is also influenced by its neighbors' neighbors. We expect to see the indirect influence as nonpairwise interactions in higher-order phase reductions.

This further highlights that any question about the importance of higher-order interactions in network dynamical systems---and thus answers to Questions~\ref{Q:Model} and~\ref{Q:Struct} in \Cref{sec:Nonadditive}---is subtle.
Even oscillator networks with \emph{additive} interactions (i.e., connections can simply be encoded by a graph) yield \emph{nonadditive} effective interactions in the (higher-order) phase reduction. 
In other words, {reducing the dynamics of a network dynamical system on a graph to an attracting lower-dimensional invariant manifold such as a torus}---a reduction in system dimension---comes at a cost of a more complicated coupling structure which encodes the nonlinearity of the vector field in a neighborhood of the invariant torus.
From this perspective it is not surprising that {the same dynamical effect, such as discontinuous synchronization transitions~\cite{Kuehn2020}, can be observed in coupled nonlinear oscillators without higher-order interactions~\cite{Calugaru2018} and phase oscillators with higher-order interactions~\cite{Skardal2019b}: The network dynamical systems may just be related to each other through a coordinate change.
This provides an opportunity to relate network dynamics with higher-order interactions (dynamical systems with specific network structure) to results from general dynamical systems theory.
}

\section{Discussion}\label{sec:discussion}

The development of a variety of tools to model and analyze high-order networks brings new opportunities to think critically and be discriminating in the types of tools that we use in different applications. It is often the case that when data is in the form of a matrix, this matrix is interpreted as the adjacency matrix or the weighted adjacency matrix of a graph.  In many cases, such an interpretation is reasonable and leads to exciting new insights in the field of application. For example, a graph, such as a connectome or interactome, is the natural mathematical structure for pairwise relational data. Other times though, a graphical interpretation is used out of necessity---there have not been the tools to analyze higher-order structures and so projecting onto the set of graphs has been the one of the few ways to perform interesting analyses. The cost of such methods however is that higher-order relationships and structure is lost, as well as some meaning.  For example, analyzing collaboration networks as graphs versus hypergraphs has been compared and contrasted in several papers with each of these articles illustrating new insights that can be gained when thinking about a collaboration network as a hypergraph~\cite{estrada2006subgraph, karwa2016discussion, lung2018hypergraph, spagnuolo2020analyzing}. Having tools for higher-order networks, such as those discussed in this review, allow such analysis to be possible, and, importantly, invites us to be more conscious about the choices that we are making when analyzing data. In terms of modeling and analysis, the development of new tools leads us to the following questions: \emph{How can we think about the trade-offs being made when making decisions about modeling? How can we balance interpretability and convenience?  And, most importantly to the quantitatively minded, is there a way to develop a guiding framework to help answer these questions?}  

While often studied separately (and by separate communities), the three aspects of higher-order networks discussed here are interlinked. 
One example is the structure--function relationship in neuroscience: How do the properties of the structural networks of neural cells and their connection relate to the dynamical properties of the network dynamical system? 
One the one hand, one can analyze the structural features of the connectivity in their own right: Whether it is extracted explicitly for small scale networks such as the stomatogastric ganglion~\cite{Harris-Warrick1992} or \emph{C.~elegans} (see \Cref{sec:Celegans}) or using techniques such as diffusion tensor imaging on the large scale, this data has been analyzed from a (higher-order) network perspective. 
Of course, these neural systems give rise to the dynamics where nonadditive higher-order effects come into play (see \Cref{sec:Neurononadd}). 
These dynamics eventually determine the function of the neural system. 
The dynamics themselves---whether empirical data from neural recordings or synthetic data from a neural model network dynamical system---can be seen as data that has been analyzed using correlation based techniques by projecting them on (higher-order) networks~\cite{Bassett2017}.

The emerging field of higher-order networks opens up exciting new directions. 
While it may sometimes appear as a patchwork of results in multiple fields, we believe there is much to be gained by cross-fertilization; this integrated review and perspective is a step in this direction. 
We are looking forward to seeing the outcomes of these research directions in the future.

\section*{Acknowledgments}
HAH is grateful for the UK Centre for TDA community, fruitful discussions with its members, with special thanks to A~Barbensi, BJ~Stolz, N~Otter, and U~Tillmann. 
CB acknowledges insightful conversations with P~Ashwin.
MTS acknowledges many interesting discussion with the participants of the Dagstuhl seminar ``Higher-Order Graph Models: From Theoretical Foundations
to Machine Learning''.

\bibliographystyle{siamplain}
\bibliography{references}
\end{document}

%% file: shared.tex
\usepackage{lipsum}
\usepackage{amsfonts,amsmath}
\usepackage{graphicx}
\usepackage{epstopdf}
\usepackage{algorithmic}
\ifpdf
  \DeclareGraphicsExtensions{.eps,.pdf,.png,.jpg}
\else
  \DeclareGraphicsExtensions{.eps}
\fi

\newsiamremark{remark}{Remark}
\newsiamremark{hypothesis}{Hypothesis}
\crefname{hypothesis}{Hypothesis}{Hypotheses}
\newsiamthm{claim}{Claim}

\headers{What are higher-order networks?}{C.~Bick, E.~Gross, H.~A.~Harrington, M.~T.~Schaub}

\title{What are higher-order networks?\thanks{Submitted to the editors DATE.
        Authors in alphabetical order.
\funding{
CB acknowledges support from the Engineering and Physical Sciences Research Council (EPSRC) through the grant EP/T013613/1.  
EG acknowledges support from the National Science Foundation (NSF) through the grant DMS-1945584.
HAH acknowledges funding from EPSRC EP/R018472/1, EP/R005125/1 and EP/T001968/1 and the Royal Society RGF$\backslash$EA$\backslash$201074 and UF150238. 
MTS received funding from the European Union’s Horizon 2020 research and innovation programme under the Marie Sklodowska-Curie grant agreement No 702410. MTS also acknowledges funding from the Ministry of Culture and Science (MKW) of the German State of North Rhine-Westphalia (``NRW Rückkehrprogramm'').
}}}

\author{
Christian~Bick\thanks{Department of Mathematics, Vrije Universiteit Amsterdam, Amsterdam, the Netherlands}
\and Elizabeth~Gross\thanks{Department of Mathematics, University of Hawai`i at M\=anoa, Honolulu, USA}
\and Heather~A.~Harrington\thanks{Mathematical Institute and Wellcome Centre for Human Genetics, University of Oxford, Oxford, UK}
\and Michael~T.~Schaub\thanks{Department of Computer Science, RWTH Aachen University}
}

\usepackage{amsopn}

\makeatletter
\newcommand*{\addFileDependency}[1]{
  \typeout{(#1)}
  \@addtofilelist{#1}
  \IfFileExists{#1}{}{\typeout{No file #1.}}
}
\makeatother

\newcommand*{\myexternaldocument}[1]{%
    \externaldocument{#1}%
    \addFileDependency{#1.tex}%
    \addFileDependency{#1.aux}%
}